\documentclass[useAMS,usenatbib]{mn2e}
\usepackage{amssymb}
\usepackage{natbib}
\usepackage{times}
\usepackage{graphics,epsfig}
\usepackage{graphicx}
\usepackage{amsmath}
\usepackage{amssymb}
\usepackage{comment}
\usepackage{supertabular}
\usepackage[flushleft]{threeparttable}

\usepackage{graphicx}
\usepackage{amsmath}
\usepackage{color}

\newcommand{\kms}{km~s$^{-1}$}
\newcommand{\arcs}{$^{\prime\prime}$}

\newcommand{\hh}{\ensuremath{\rm H_{2}}}

\newcommand{\mspcc}{\ensuremath{\rm M_{\odot} pc^{-3} }}
\newcommand{\HI}{H{\sc i}}

\title[\HI~scale height in spiral galaxies]{\HI~scale height in spiral galaxies}
\author [N. N. Patra]{	Narendra Nath Patra$^{1}$ \thanks {E-mail: narendra@rri.res.in} \\
	$^{1}$ Raman Research Institute, C. V. Raman Avenue, Sadashivanagar, Bengaluru 560080, India\\
}
\date {}
\begin{document}
\maketitle
%\pagerange{\pageref{firstpage}--\pageref{lastpage}} \pubyear{}
%\label{firstpage}

\begin{abstract}

We model the galactic discs of seven nearby large spiral galaxies as three-component systems consist of stars, molecular gas, and atomic gas in vertical hydrostatic equilibrium. We set up the corresponding joint Poisson-Boltzmann equation and solve it numerically to estimate the three-dimensional distribution of \HI~in these galaxies. While solving the Poisson-Boltzmann equation, we do not consider a constant \HI~velocity dispersion ($\sigma_{\rm HI}$); rather, we develop an iterative method to self-consistently estimate the $\sigma_{\rm HI}$ profile in a galaxy by using the observed second-moment profile of the \HI~spectral cube. Using the density solutions, we determine the \HI~vertical scale height in our galaxies. We find that the \HI~discs flare in a linear fashion as a function of radius. \HI~scale height in our galaxies is found to vary between a few hundred parsecs at the center to $\sim 1-2$ kpc at the outskirts. We estimate the axial ratio of the \HI~discs in our sample galaxies and find a median ratio of 0.1, which is much lower than what is found for dwarf galaxies, indicating much thinner \HI~discs in spiral galaxies. Very low axial ratios in three of our sample galaxies (NGC 5055, NGC 6946, and NGC 7331) suggest them to be potential superthin galaxies. Using the \HI~distribution and the \HI~hole sizes in NGC 6946, we find that most of the \HI~holes in this galaxy are broken out into the circumgalactic medium and this breaking out is more effective in the inner radii as compared to the outer radii.

\end{abstract}


\section{Introduction}

The atomic hydrogen (\HI) in galaxies is one of the most critical components which directly influences the physical, chemical, and dynamical evolution of a galaxy. The atomic gas acts as a long-term fuel for star formation and found to correlate with the current star formation in both spiral and dwarf galaxies \citep{kennicutt98b,bigiel08,roychowdhury09,patra16,leroy09b,roychowdhury15,bacchini19a, bacchini19b}. Though the stars born out of molecular beds, the molecular gas forms out of the Cold Neutral Medium (CNM) phase of the interstellar medium (ISM). The connection between the ISM phases and the star formation requires the knowledge of the density distribution of the \HI~in galaxies, which might play a pivotal role in deciding different phase transitions and inducing instabilities \citep{toomre64,madore77,burton87,wolfire95b,alves07}. In fact, in a recent study, \citet{bacchini19a} have shown that a volumetric star-formation law shows a much lower scatter than the traditional star formations laws with surface densities. This, in turn, indicates that the volume density distribution of the gas in galaxies is crucial in understanding the origin of star formation and hence, the star formation laws.

\HI~also plays a vital role in understanding the dynamical evolution of galaxies. For example, \HI~extends to much larger radii as compared to optical discs and serves as a tracer of the underlying dark matter halo potential at these regions. Such as, to date, the \HI~rotation curve is the most robust indicator of the presence of dark matter in galaxies, and it is routinely used to investigate the properties of the dark matter halos \citep{bosma78,bosma81,vanalbada85,begeman87,deblok08,oh15,iorio17}. Moreover, the distribution of \HI~is an excellent tracer of the gravitational potential in the vertical direction as well. For example, in the Galaxy, it is observed that the \HI~flares as a function of the galactocentric radius. In that sense, the \HI~can trace the variation in the gravitational potential (and hence the underlying dark matter density) in the vertical direction \citep{banerjee08}, especially on the outskirts of a galaxy where the dark matter dominates.

\begin{table*}
\caption{Basic properties of our sample galaxies}
\begin{threeparttable}
\begin{tabular}{lccccccccc}
\hline
Name & RA (J2000) & DEC (J2000)  & Dist &  $\rm M_B$  & $\rm r_{25}$  & $\rm i$ & PA & $\rm \log M_{HI}$ & SFR \\
    & (h m s)    & ($^o$ $^\prime$ $^{\prime \prime}$) & (Mpc) & (mag) & ($^\prime$) & ($^o$) & ($^o$) & ($\rm M_{\rm \odot}$) & ($\rm M_{\rm \odot} \thinspace yr^{-1}$) \\
(1)     &      (2)      & (3)  &      (4)     &   (5)     &   (6)     & (7)  & (8) & (9)  & (10)\\
\hline
NGC 925 & 02 27 16.5 & +33 34 44 & 9.2 & -20.04 &  5.3 & 66 & 287 & 9.66 & 1.09 \\ 
NGC 2841 & 09 22 02.6 & +50 58 35 & 14.1 & -21.21 &  5.3 & 74 & 154 & 9.93 & 0.20 \\ 
NGC 2976 & 09 47 15.3 & +67 55 00 & 3.6 & -17.78 &  3.6 & 65 & 335 & 8.13 & 0.10 \\ 
NGC 3198 & 10 19 55.0 & +45 32 59 & 13.8 & -20.75 &  3.2 & 72 & 215 & 10.01 & 0.85 \\ 
NGC 5055 & 13 15 49.2 & +42 01 45 & 10.1 & -21.12 &  5.9 & 59 & 102 & 9.96 & 2.42 \\ 
NGC 6946 & 20 34 52.2 & +60 09 14 & 5.9 & -20.61 &  5.7 & 33 & 243 & 9.62 & 4.76 \\ 
NGC 7331 & 22 37 04.1 & +34 24 57 & 14.7 & -21.67 &  4.6 & 76 & 168 & 9.96 & 4.20 \\ 
\hline
\end{tabular}
\begin{tablenotes}
\item NOTE: The data presented in this table is compiled from \citet{walter08}.
\end{tablenotes}
\label{tab:samp}
\end{threeparttable}
\end{table*}

The three-dimensional distribution of \HI~in galaxies is also crucial to many other physical properties of galaxies. For example, the thickness of the \HI~discs in galaxies influences the effective cross-section to the background quasars, which produces the absorption spectra. Thicker \HI~discs in a particular type of galaxies would also increase their incident rates as a function of redshift as well \citep[see, e.g., ][]{zwaan05b,patra13,roy13c}. Further, the thickness of the \HI~layer in a galaxy decides the size distribution of the \HI~holes originated due to supernovae or stellar feedback. Consequently, it determines if the \HI~shell would break into the circumgalactic medium (CGM) polluting it with metal-enriched gas or not \citep{maclow99,silich01}. However, despite its tremendous importance, a direct determination of the three-dimensional distribution of the \HI~is difficult in external galaxies due to the line-of-sight integration effect. For the Galaxy on the other hand, determination of the \HI~distribution is possible with direct observations assisted with minimal modeling \citep{grabelsky87,wouterloot90,nakanishi03,rathborne09b,kalberla08,roman-duval10,roman-duval16,sofue17,marasco17}. For external galaxies, using observation only, a direct estimation of the vertical \HI~distribution (and hence the \HI~scale height (defined as the Half Width at Half Maxima of the vertical density distribution)) without modelling is only possible if the galaxy has a constant vertical scale height and, viewed at an edge-on orientation. However, for almost no galaxies, the \HI~vertical scale height is constant as a function of radius. Hence, detailed modeling of the \HI~disc is essential to estimate the three-dimensional distribution of \HI~in a galaxy.

In fact, a large number of previous studies has estimated the three-dimensional distribution of stars and gas in a number of galaxies using detailed hydrostatic modeling considering galactic discs to be multi-component isothermal systems \citep{woolley57,vanderkruit81a,vanderkruit81b,vanderkruit81c,bahcall84a,bahcall84b,bahcall84c,bahcall91,romeo92,olling95,narayan02b,abramova08,banerjee11b,bacchini19a,bacchini19b}. In early studies, \citet{bahcall84a,bahcall84b} assumed the stellar disc of the Galaxy to be an isothermal system in vertical hydrostatic equilibrium and solved the combined Boltzman-Poissons equation. They estimated the vertical stellar mass distribution and consequently measured the ratio of the halo to disc mass densities. They also used this method to determine the relationship between the total surface density of mass, observed luminosity scale height, and vertical velocity dispersion \citep{bahcall84c}. Later, \citet{romeo92} solved the hydrostatic equation assuming galactic discs to be two-component systems (stars+cold gas) to investigate the self-regulation mechanisms in determining the dynamics and structure of early normal spiral galaxies. The hydrostatic condition has further been used to investigate the shape of the dark matter halos in galaxies by modeling the vertical \HI~distribution \citep[see, e.g.,][]{olling95,banerjee08}.

In this work, we model the baryonic discs of seven nearby large spiral galaxies from The \HI~Nearby Galaxy Survey (THINGS, \citet{walter08}) to estimate the three-dimensional distribution of \HI~in them. We consider the galactic discs to be three-component systems consist of stars, molecular gas, and atomic gas in vertical hydrostatic equilibrium under their mutual gravity and the dark matter halo. We further set up the joint Poisson-Boltzmann equation of hydrostatic equilibrium and solve it numerically to obtain a detailed distribution of the \HI~in these galaxies. 

To solve the Poisson-Boltzmann equation numerically, we adopt a similar strategy as used by \citet{narayan02b} and subsequent studies \citep{banerjee08,banerjee10,patra14,patra19b,patra18a,sarkar18,sarkar19,patra20a,patra20b}. However, most of these previous studies considered the vertical velocity dispersion of \HI~($\sigma_{\rm HI}$) to be constant for simplicity or lack of observation. Recently, in rigorous studies, \citet{bacchini19a,bacchini19b} used a very similar technique to obtain a three-dimensional distribution of \HI,\hh~and star formation rate in a sample of twelve nearby star-forming galaxies and the Milky Way. Unlike previous studies, they extracted the vertical velocity dispersion of the gas discs as a function of the radius by performing detailed 3D tilted ring modeling of the kinematic data using $^{3D}$Barolo \citep{diteodoro15}. However, they did not solve the hydrostatic equation fully self-consistently and ignored the gravity of the \hh~disc while solving for \HI. Whereas, in this study, we solve the hydrostatic equilibrium equation self-consistently incorporating the gravity of all the disc components. We also do not assume a constant $\sigma_{\rm HI}$; rather, we develop an iterative method by which we estimate the $\sigma_{\rm HI}$ profile in a galaxy from the observed second-moment profile (MOM2) of the \HI~spectral cube. In this approach, simultaneously, both the \HI~distribution and the $\sigma_{\rm HI}$ can be determined self-consistently in a galaxy.

\section{Sample}

We select seven nearby spiral galaxies from The \HI~Nearby Galaxy Survey (THINGS, \citet{walter08}) for which all the necessary data are available publicly. As part of the THINGS survey, a total of 34 galaxies were observed in \HI~with the Very Large Array (VLA) with a high spatial resolution of $\sim$ 6\arcs. The sample was carefully chosen, such as to be part of a larger multi-wavelength campaign. For example, THINGS galaxies are part of the {\it Spitzer} Infrared Nearby Galaxy Survey (SINGS, \citet{kennicutt03}), which enables robust estimation of the stellar masses in these galaxies. They have the FUV data from the GALEX observations, which is useful to calculate the recent star formation in these galaxies. Moreover, the HERA CO Line Extra-galactic Survey (HERACLES, \citet{leroy09b}) also used the THINGS samples as the base to observe 18 large spiral galaxies with the 30-m IRAM telescope and mapped molecular gas in these galaxies.

Solving the Poisson-Boltzmann equation requires several essential input parameters. For every galaxy, one needs the surface density profiles of the disc components, which are primarily responsible for gravity. Here it should be mentioned that several galaxies in the THINGS sample did not have publicly available molecular surface density profiles. As molecular gas is one of the crucial baryonic components which can significantly influence the dynamics, we exclude these galaxies from our sample. \HI~rotation curve and the mass-model (dark matter distribution) of a galaxy are the other essential inputs to the hydrostatic equation. However, due to limitations in the observing parameters (e.g., inclination, data quality etc.), mass-modeling could not be performed in some of the THINGS galaxies \citep[see][for more details]{deblok08}. We exclude these galaxies as well from our sample. This leads to a total of eight galaxies for which all the necessary parameters are available to solve the hydrostatic equation. However, for one galaxy, NGC 3521, the \HI~disc shows strong warp at the outer disc \citep[see, e.g., Fig. 77 of][]{deblok08}. This results in a varying inclination of the annular ring at a particular radius and can overestimate the observed $\sigma_{\rm HI}$ significantly. Hence, we exclude this galaxy from our sample which leads to a total seven galaxies in our sample for which we solve the hydrostatic equilibrium equation. 

In Tab.~\ref{tab:samp}, we present a few basic properties of our sample galaxies. Column (1) quotes the name of the galaxies, column (2) and (3) present the coordinates (RA and DEC) of the galaxies. In column (4) and (5), we show the distances to the galaxies and the absolute blue band magnitude (corrected for the extinction of the Milky Way and the galaxy itself), respectively. Column (6) denotes the optical radius ($r_{\rm 25}$), whereas column (7) and (8) present the average inclination and the position angle of the \HI~discs as obtained by a tilted-ring fitting of the \HI~velocity field (as obtained by \citet{deblok08}). Finally, column (9) shows the \HI~mass of the galaxies, whereas column (10) depicts the global star formation rate. The data presented in this table are compiled from \citet{walter08}.

\section{Modeling the galactic discs}

\begin{figure*}
\begin{center}
\begin{tabular}{c}
\resizebox{1.\textwidth}{!}{\includegraphics{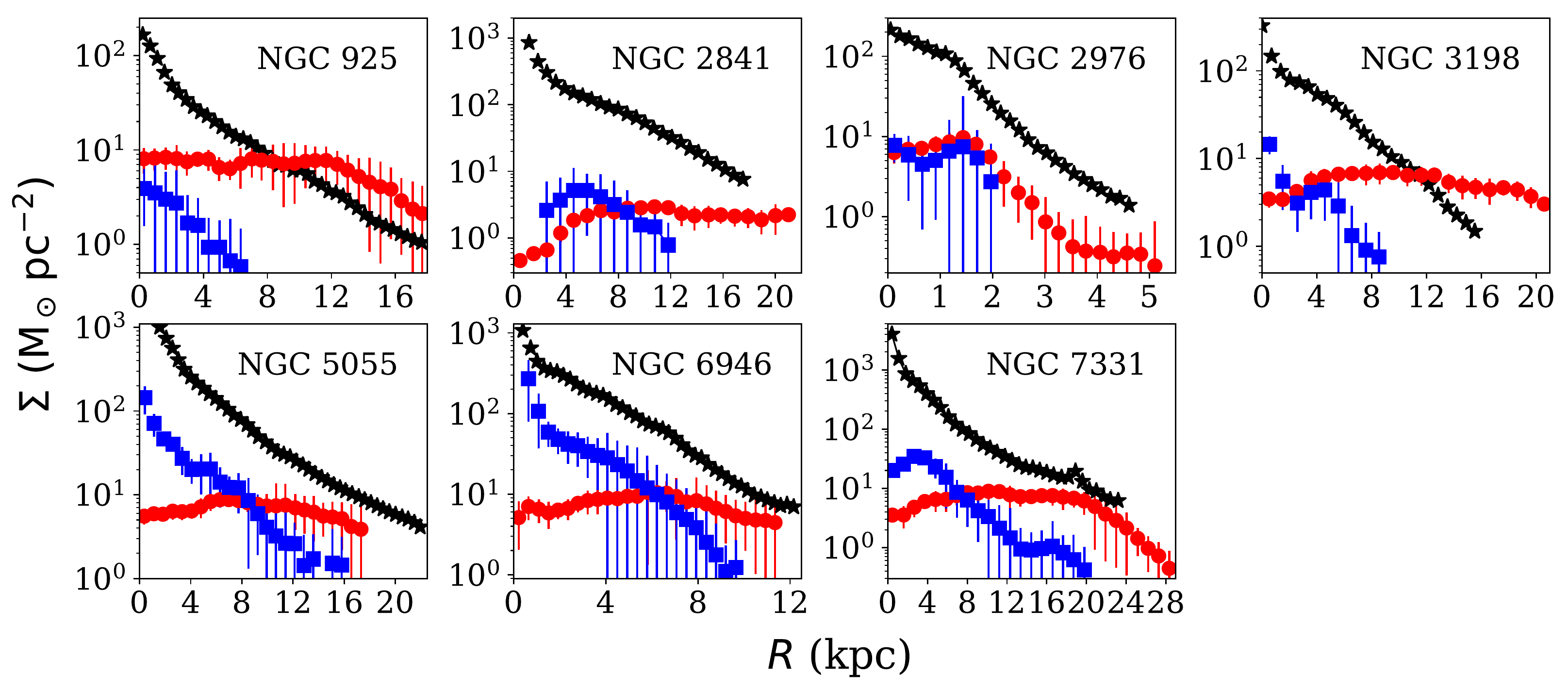}}
\end{tabular}
\end{center}
\caption{The surface density profiles of our sample galaxies. The black stars represent the stellar surface density profiles, whereas the red circles with error bars depict the \HI~surface density profiles. The molecular surface densities are shown by the blue squares with error bars. Each panel represents different galaxies, as quoted at the top right corner of the panel. The stellar surface density data was taken from \citet{leroy08}, whereas the \HI~and the molecular surface density profiles are obtained from \citet{schruba11}.}
\label{sden}
\end{figure*}

We model our sample galaxies as three-component systems consisting of stellar, molecular, and atomic discs. We assume that all these baryonic discs are in vertical hydrostatic equilibrium under their mutual gravity in the external force field of the dark matter halo. The dark matter potential is considered here to be fixed as determined by the observed rotation curve (or mass model). In hydrostatic equilibrium, the gravity will be balanced by the pressure. The gravity on any elemental volume (in any disc component) will be resulted due to all the three components, whereas, the pressure would be determined only by the velocity dispersion of an individual component. For simplicity, we assume all the three baryonic discs are concentric and coplanar with the centers coinciding with that of the dark matter halo. In that case, the Poisson's equation of hydrostatic equilibrium at any radius can be written in cylindrical polar coordinate as 

\begin{equation}
\frac{1}{R} \frac{\partial }{\partial R} \left( R \frac{\partial \Phi_{\rm tot}}{\partial R} \right) + \frac{\partial^2 \Phi_{\rm tot}}{\partial z^2} = 4 \pi G \left( \sum_{i=1}^{3} \rho_{\rm i} + \rho_{\rm h} \right)
\label{eq1}
\end{equation}

\noindent where $\Phi_{\rm tot}$ is the total potential due to the gravity of the baryonic discs and the dark matter halo. $\rho_{\rm i}$ represents the volume density of different disc components, where $i$ runs for stars, \HI, and molecular gas. $\rho_{\rm h}$ denotes the density of the dark matter halo.

In hydrostatic equilibrium, the gradient in the potential in the vertical direction would be balanced by the gradient in pressure. This can be represented by the Boltzmann equation as 

\begin{equation}
\frac{\partial }{\partial z} \left(\rho_{\rm i} {\langle {\sigma}_{\rm z}^2 \rangle}_{\rm i} \right) + \rho_{\rm i} \frac{\partial \Phi_{\rm tot}}{\partial z} = 0
\label{eq2}
\end{equation}

\noindent where $\sigma_{\rm z}$ represents the velocity dispersion of different baryonic discs and is a measure of the vertical pressure. Eq.~\ref{eq2} can be used to simplify Eq.~\ref{eq1} as

\begin{equation}
\begin{split}
{\langle {\sigma}_{\rm z}^2 \rangle}_{\rm i} \frac{\partial}{\partial z} \left( \frac{1}{\rho_{\rm i}} \frac{\partial \rho_{\rm i}}{\partial z} \right) &= \\ 
&-4 \pi G \left( \rho_{\rm s} + \rho_{\rm HI} + \rho_{\rm H2} + \rho_{\rm h} \right)\\ 
&+ \frac{1}{R} \frac{\partial}{\partial R} \left( R \frac{\partial \Phi_{\rm tot}}{\partial R} \right)
\end{split}
\label{eq3}
\end{equation}

\noindent where, $\rho_{\rm s}$, $\rho_{\rm H2}$ and $\rho_{\rm HI}$ are the volume densities of stars, molecular gas, and atomic gas, respectively. The rightmost term in the RHS of Eq.~\ref{eq3} originates due to the gradient of the potential in the radial direction. This can be estimated using the observed rotation curve. The \HI~rotation curve is expected to trace the potential of a galaxy in the midplane. In that sense, the radial term can be expressed as  

\begin{equation}
{\left( R \frac{\partial \Phi_{\rm tot}}{\partial R} \right)}_{\rm R,z} = {(v_{\rm rot}^2)}_{R,z}
\label{eq4}
\end{equation}

\noindent where the $v_{\rm rot}$ (measured using the Doppler shift of the \HI-21cm line) is implicitly taken to be equal to the circular speed. Though the above equation is strictly valid for $z=0$, we assume that the rotation curve does not change considerably as a function of $z$, and it traces the potential above the midplane reasonably well. With this assumption, Eq.~\ref{eq3} can further be simplified using Eq.~\ref{eq4} to rewrite the hydrostatic equilibrium equation as

\begin{equation}
\begin{split}
{\langle {\sigma}_{\rm z}^2 \rangle}_{\rm i} \frac{\partial}{\partial z} \left( \frac{1}{\rho_{\rm i}} \frac{\partial \rho_{\rm i}}{\partial z} \right) &= \\
&-4 \pi G \left( \rho_{\rm s} + \rho_{\rm HI} + \rho_{\rm H2} + \rho_{\rm h} \right)\\ 
&+ \frac{1}{R} \frac{\partial}{\partial R} \left( v_{\rm rot}^2 \right)
\end{split}
\label{eq5}
\end{equation}

Eq.~\ref{eq5} represents three second-order partial coupled ordinary differential equations in $\rho_{\rm s} (z)$, $\rho_{\rm H2} (z)$ and $\rho_{\rm HI} (z)$. The solutions of these equations at any radius would provide the density distribution, $\rho_{\rm i}$, as a function of $z$. To produce a complete three-dimensional density distribution of any disc component, one needs to solve these equations at all radii.

\subsection{Input parameters}

Several input parameters are required to solve Eq.~\ref{eq5} in a self-consistent manner. As the baryonic discs provide a significant amount of gravity to the hydrostatic equation, the surface density profiles play an essential input to Eq.~\ref{eq5}. In Fig.~\ref{sden}, we show the surface density profiles of our sample galaxies. As can be seen from the figure, the stellar discs dominate the surface densities in all our sample galaxies at the inner parts whereas, the \HI~dominates the outskirts. It can also be seen that the molecular discs in our galaxies are considerably shorter as compared to the stellar and the \HI~discs. Because of this reason, at all radii, we cannot consider our galaxies as three-component systems (stars+\HI+\hh). We, therefore, consider our galaxies to be three-component systems till the edge of the molecular discs; beyond that, we consider them to be two-component systems with stars and \HI.

Another input to the hydrostatic equation is the density distribution of the dark matter halo. Particularly for \HI~distribution, the dark matter can play a significant role as the \HI~extends to regions (both in radial and the vertical direction) where dark matter dominates the gravity. We use the mass-models of our sample galaxies from \citet{deblok08} and obtain the dark matter halo parameters. The mass-modeling of our sample galaxies were done using both an isothermal and an NFW halo. The NFW dark matter density profile can be given as  

\begin{equation}
\label{nfw}
\rho_{h} (R) = \frac {\rho_{\rm 0}}{\frac{R}{R_{\rm s}} \left( 1 + \frac{R}{R_{\rm s}}\right)^2}
\end{equation}

\noindent whereas, the pseudo-isothermal profile (ISO) can be given as

\begin{equation}
\label{eq_iso}
\rho_{\rm h}(R) = \frac {\rho_{\rm 0}}{1 + \left(\frac{R}{R_{\rm s}}\right)^2}
\end{equation}

\noindent where $\rho_{\rm 0}$ is the characteristic density, and $R_{\rm s}$ is the characteristic radius. These two parameters completely describe a spherically symmetric dark matter halo. As discussed in detail in \citet{deblok08} (see their Table 3 and Table 4 for more details), no particular dark matter density profile (ISO or NFW) describes the rotation curves of our galaxies systematically better than the other one. Hence, depending on the galaxy, we choose a dark matter halo profile that better describes its rotation curve (based on the reduced $\chi^2$ values of the fitted rotation curves). However, it should be noted that this choice does not influence our results significantly as both the dark matter profiles explain the rotation curves reasonably well within the observational uncertainties. In Tab.~\ref{tab:dmpar}, we present the dark matter halo parameters we adopt to solve Eq.~\ref{eq5} for our sample galaxies.

\begin{table}
\caption{Dark matter halo parameters}
\begin{center}
\begin{tabular}{lccc}
\hline
Galaxy & Profile & $R_s$ & $\rho_0$ \\
     &         & (kpc) & ($\times 10^{-3} $ \mspcc)\\
\hline
NGC 925   & ISO & 9.67  & 5.90\\
NGC 2841  & NFW & 20.55 & 12.40\\
NGC 2976  & ISO & 5.09  & 35.50\\
NGC 3198  & ISO & 2.71  & 47.50\\
NGC 5055  & ISO & 11.73 & 4.80\\
NGC 6946  & ISO & 3.62  & 45.70\\
NGC 7331  & NFW & 60.20 & 1.05\\
\hline
\end{tabular}
\end{center}
\label{tab:dmpar}
\end{table}

\begin{figure*}
\begin{center}
\begin{tabular}{c}
\resizebox{1.\textwidth}{!}{\includegraphics{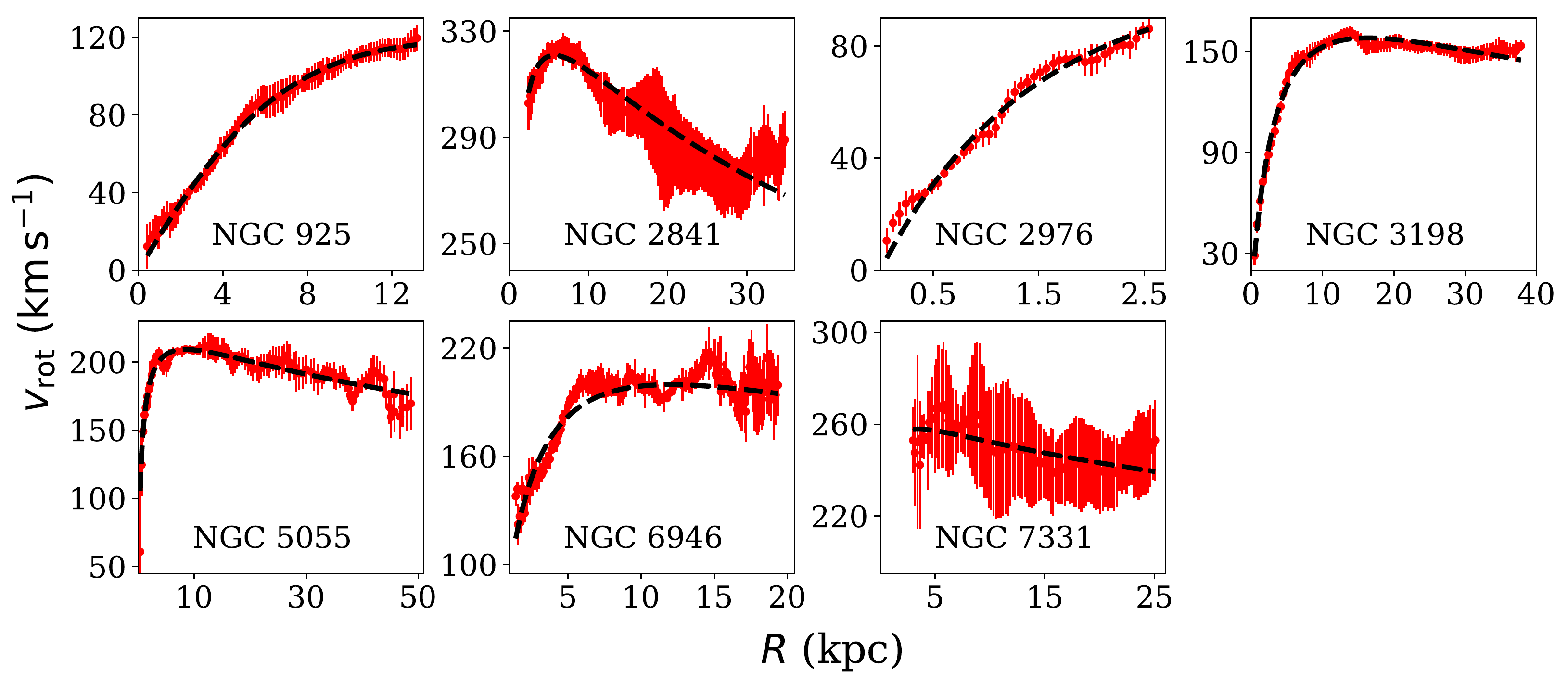}}
\end{tabular}
\end{center}
\caption{The rotation curves of our sample galaxies as obtained from \citet{deblok08}. The solid red circles with error bars represent the rotation curves as obtained by a tilted-ring fitting of the \HI~velocity fields. The black dashed lines in each panel represent a Brandt profile fit to the rotation curves. See the text for more details.}
\label{rotcur}
\end{figure*}

The observed rotation curves are another input parameter required to evaluate the radial term in Eq.~\ref{eq5}. We obtain the rotation curves of our sample galaxies from \citet{deblok08}. In Fig.~\ref{rotcur}, we show the rotation curves of our sample galaxies. As can be seen from Eq.~\ref{eq5}, the radial term is computed by calculating the first derivative of the rotation curve's square. In that case, any sudden change/jump in the rotation curve (could be due to measurement/modeling limitations), might lead to an unphysical value of the derivative and hence diverge the solutions. To overcome this shortcoming, we fit the rotation curves with a commonly used Brandt profile \citep{brandt60}. A Brandt profile can be given as
\begin{equation}
v_{\rm rot} (R) = \frac{V_{\rm max}\left(R/R_{\rm max} \right)}{\left(1/3 + 2/3 \left(\frac{R}{R_{\rm max}}\right)^n\right)^{3/2n}}
\end{equation}

\noindent where $v_{rot}$ is the rotation velocity, $V_{max}$ is the maximum attained velocity and $R_{max}$ is the radius at which the $V_{max}$ is attained. $n$ is the index, which decides how fast or slow the rotation curve rises to $V_{max}$. In Tab.~\ref{tab:rotcur}, we present the fit parameters of the rotation curves. We use these parameters to calculate the radial term, which is now free from any sudden jump/variation.

\begin{table}
\caption{Best fit parameters of the rotation curves}
\begin{center}
\begin{tabular}{lccc}
\hline
Galaxy & $V_{\rm max}$ & $R_{\rm max}$ & $n$ \\
     &(\kms) & (kpc) & \\
\hline
NGC 925   & 118.0$\pm$1.4 & 16.4$\pm$0.9 & 1.80$\pm$0.10\\
NGC 2841  & 320.9$\pm$0.5 & 5.7$\pm$0.2  & 0.36$\pm$0.02\\
NGC 2976  & 103.5$\pm$13.6& 6.9$\pm$3.0  & 1.05$\pm$0.28\\
NGC 3198  & 158.1$\pm$0.7 & 16.4$\pm$0.3 & 0.82$\pm$0.03\\
NGC 5055  & 209.1$\pm$0.6 & 8.5$\pm$0.2  & 0.36$\pm$0.01\\
NGC 6946  & 199.7$\pm$0.5 & 12.0$\pm$0.6 & 0.69$\pm$0.05\\
NGC 7331  & 257.8$\pm$1.7 & 3.5$\pm$1.3  & 0.12$\pm$0.04\\
\hline
\end{tabular}
\end{center}
\label{tab:rotcur}
\end{table}

The vertical velocity dispersions of the baryonic disc components are one of the most critical inputs to Eq.~\ref{eq5}. The velocity dispersion alone decides the resistive force balancing the total gravity due to all the components and the dark matter halo on an elemental volume. Moreover, as gravity is coupled, the influence of a particular component's gravity is shared by all the disc components. On the other hand, the stability of a disc component (stars, molecular gas, or \HI) against this combined gravity is solely controlled by its velocity dispersion. In this regard, the velocity dispersion can influence the density distribution significantly, and hence, a precise measurement of the same is essential.

Direct measurement of the stellar velocity dispersion is tough even with present-day optical telescopes. Because a direct measurement is not available, we calculate the stellar velocity dispersion analytically using the scale length of the optical disc, $R_{\rm D}$, and the stellar surface density as described in \citet{leroy08} (given in Appendix B in their paper). In this calculation, the stellar disc is assumed to be a single-component system in hydrostatic equilibrium. As this calculation ignores gas-gravity, the calculated stellar velocity dispersion is slightly underestimated. 

Unlike the stellar velocity dispersion, the gas velocity dispersion can be estimated relatively easily using spectroscopic observations by radio telescopes. In the context of the molecular velocity dispersion in the Galaxy, early observations by \citet{stark84} reveal that low-mass molecular clouds have a velocity dispersion of $\sim 9$ \kms, whereas, the high-mass clouds have a relatively lower velocity dispersion of $\sim 6.6$ \kms. It should be mentioned here that the gas velocity dispersion refers to the dispersion due to the relative turbulent motions between the gas clouds. The thermal velocity dispersion (due to the kinetic temperature) within a molecular cloud is expected to be much less ($\sim$ 0.1 \kms). In a more recent study, \citet{calduprimo13} used the data from the THINGS and the HERACLES survey for 12 nearby spiral galaxies to estimate the velocity dispersion of \HI~and CO. Using a spectral stacking method (for improved SNR), they found $\sigma_{\rm HI}/\sigma_{\rm CO} = 1.0 \pm 0.2$ with a median $\sigma_{\rm HI} = 11.9 \pm 3.1$ \kms. Later, \citet{mogotsi16} used the same sample to investigate the individual high SNR spectra and found a $\sigma_{\rm HI}/\sigma_{\rm CO} = 1.4 \pm 0.2$. They concluded that the molecular disc has a diffuse component with somewhat higher velocity dispersion, which is not detected in individual spectra, but only recovered when multiple spectra are stacked. This indicates that the dominant component in the molecular disc, which is detected in individual spectra, has a velocity dispersion of $\sim$ 7 \kms. Hence, for our sample galaxies, we adopt a velocity dispersion of 7 \kms~for their molecular discs. However, in a recent study, \citet{marasco17} have found a somewhat lower velocity dispersion ($4.4 \pm 1.2$ \kms) of the molecular gas in the Galaxy inside the solar cycle. Nevertheless, the velocity dispersion of the stellar discs or the molecular discs does not influence the density distribution of \HI~considerably \cite[see, e.g.,][]{banerjee11b,patra19b}. Hence, our adopted velocity dispersions for the stellar and the molecular discs are adequate for solving Eq.~\ref{eq5}.

\begin{figure}
\begin{center}
\begin{tabular}{c}
\resizebox{.47\textwidth}{!}{\includegraphics{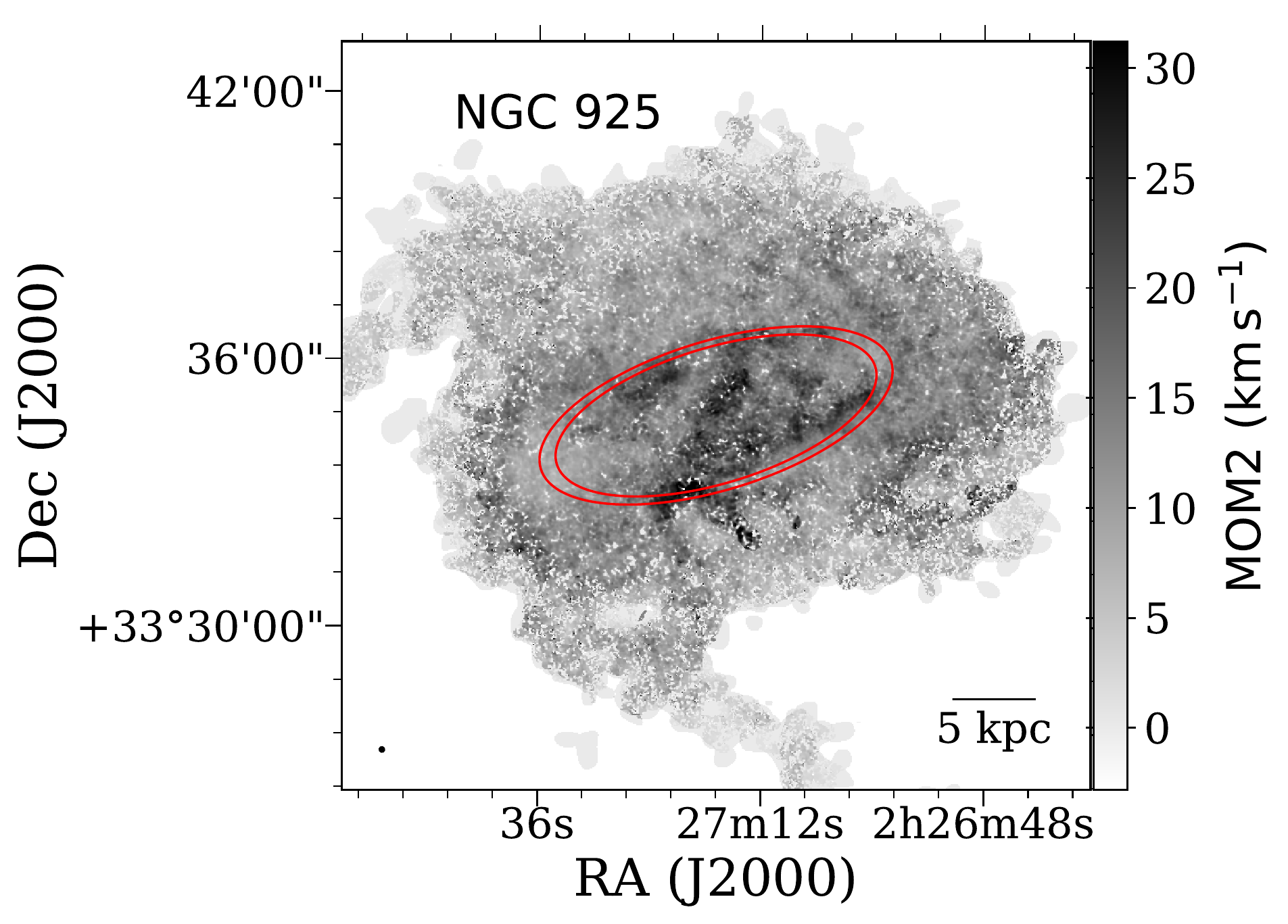}}
\end{tabular}
\end{center}
\caption{The MOM2 map of a representative galaxy, NGC 925 in our sample. The grayscale indicates the MOM2 distribution in the galaxy. We extract the MOM2 profile of our galaxies by averaging the MOM2 values in radial bins. The red ellipse represents a radial bin at 10 kpc distance from the center. The solid circle at the bottom left corner represents the observing beam. Whereas, the scale bar at the bottom right represents a linear scale of 5 kpc at the distance of the galaxy, 9.2 Mpc.}
\label{sampl_mom2}
\end{figure}

Unlike the stellar and the molecular velocity dispersions, the \HI~velocity dispersion plays a crucial role in deciding the density distribution in an \HI~disc. Hence, careful estimation of the same is essential while solving Eq.~\ref{eq5}. Early spectroscopic studies with a low spatial resolution reveal an \HI~velocity dispersion ($\sigma_{\rm HI}$) in the range 6-13 \kms~in nearby galaxies \citep{shostak84,vanderkruit84,kamphuis93}. However, more recent studies used high spatial resolution spectral cubes to estimate $\sigma_{\rm HI}$ in nearby large spiral galaxies. For example, in an extensive analysis of the second-moment (MOM2) of the \HI~spectral cubes from the THINGS data, \citet{tamburro09} found a significant variation in $\sigma_{\rm HI}$ within a galaxy as well as across the galaxies. They found a mean $\sigma_{\rm HI}$ of $\sim 10$ \kms~at the optical radius, $r_{\rm 25}$. These results suggest that the $\sigma_{\rm HI}$ might have a significant variation within and across our sample galaxies. 

To estimate the \HI~velocity dispersion, we use the publicly available moment maps of our sample galaxies as obtained in the THINGS survey\footnote{https://www2.mpia-hd.mpg.de/THINGS/Data.html}. We refer the readers to \citet{walter08} for a detailed description of generating the moment maps from the observed \HI~spectral cubes. In Fig.~\ref{sampl_mom2}, we show an example MOM2 map of a representative galaxy, NGC 925, from our sample. As can be seen from the figure, MOM2 in NGC 925 shows considerable variation across the galaxy. To capture this variation as a function of radius and to obtain a MOM2 profile, we construct a number of annuli as a function of the radius with a width of $\sim 2$ times the spatial resolution. In Fig.~\ref{sampl_mom2}, we show one such annulus with red ellipses at a radius of 10 kpc. We generate a MOM2 profile by averaging all the MOM2 values within an annulus as a function of the radius. In Fig.~\ref{sig_opt}, the first panel shows the resulting MOM2 profile (solid red circles with error bars) for NGC 925.

It should be noted that due to the low SNR, some of the pixels in Fig.~\ref{sampl_mom2} are blanked while producing the moment maps (see, \citet{walter08} for more details). We exclude these pixels while estimating the average MOM2 value within an annulus. Nevertheless, this has no considerable effect on our estimation of MOM2 profiles as the fraction of these blanked pixels is small. However, a low SNR (especially at the outskirts of the galaxies) can have a non-negligible effect on the determined MOM2 profile of a galaxy. At low SNRs, the MOM2 only traces the peaks of the \HI~profiles and hence, artificially underestimates the actual $\sigma_{\rm HI}$. We find that the MOM2 could underestimate the $\sigma_{\rm HI}$ by $\sim 10\%$ in a region where SNR is $\lesssim 3$ in MOM0 map. This value reduces to a few percent when the SNR increases to $\gtrsim 5$. For our galaxies, we found the observed SNR to be $\gtrsim 3$ in the regions of our interests. Hence, we include a conservative error of $10\%$ in our estimated MOM2 values. Further, to account for the variation of the MOM2 within an annulus, we add the normalized (by the square root of the number of independent beams within an annulus) standard deviation in quadrature to the error bars. As can be seen from Fig.~\ref{sig_opt}, the MOM2 profiles for our sample galaxies do not remain constant, rather vary considerably as a function of radius. 

However, MOM2 does not represent the intrinsic velocity dispersion; instead, it is the intensity weighted $\sigma_{\rm HI}$ along a line-of-sight. For galaxies where the beam-smearing of line-of-sight \HI~spectra is significant, the MOM2 can considerably differ from $\sigma_{\rm HI}$. Especially in the central regions of galaxies, where the gradient of the rotation curve is high. Three primary factors mainly influence the smearing of the rotational velocity into the \HI~spectral width. Firstly, the slope of the rotation curve, secondly the spatial resolution, and thirdly the inclination of the \HI~disc. The thickness of the \HI~disc also can contribute to the smearing, albeit on a minor level. Due to these smearing effects, the MOM2 would always be an overestimate of the intrinsic $\sigma_{\rm HI}$ profile. Hence, using the MOM2 profile in Eq.~\ref{eq5} as input $\sigma_{\rm HI}$ is not suitable. Instead, we use an iterative method to estimate the intrinsic $\sigma_{\rm HI}$ profile in a self-consistent manner using the observed MOM2 profile. The details of this method are explained in the next section.

\section{Solving the hydrostatic equation}

Eq.~\ref{eq5} is a second-order ordinary coupled partial differential equation, which cannot be solved analytically even for a two-component system. Hence, we adopt a numerical approach to solve it. We use an eighth-order Runge-Kutta method as implemented in the {\tt python} package {\tt scipy} to integrate the differential equation and obtain the density solutions. To integrate Eq.~\ref{eq5}, we need to know the {\it `initial condition'} or the values of $\frac {\partial \rho_{\rm i}} {\partial z}$ and $\rho_{\rm i}$ at $z=0$, from where we start the integration. We can adopt 

\begin{equation}
\left( \rho_{\rm i} \right)_{\rm z = 0} = \rho_{\rm i,0} \ \ \ \ {\rm and} \ \ \ \left(\frac{\partial \rho_{\rm i}} {\partial z}\right)_{\rm z=0} = 0
\label{init_cond}
\end{equation}

\noindent Due to vertical symmetry, the density at the midplane is expected to be maximum, leading to the second condition. However, for the first condition, one needs to know the midplane density at every radius to initiate the integration. However, this midplane density is not a directly measurable quantity in galaxies (except for the Galaxy, \citep[see, e.g., ][]{kalberla08,grabelsky87,grabelsky88,wouterloot90,lopez-corredoira14,wang18}). We estimate this midplane density at every radius using the knowledge of the surface densities at that radius.

Adopting a similar strategy as used by many previous authors \citep{narayan02b,banerjee11b,patra14,patra18a,patra19b}, we use an iterative approach to find the midplane density using the knowledge of the observed surface density. For solving at a particular radius for a particular component, say, stars, we first assume a trial midplane density $\rho_{\rm i,0,t}$, and solve Eq.~\ref{eq5}. This will result in a trial solution $\rho_{\rm i,t}(z)$. This solution is not the correct solution as we chose the $\rho_{\rm i,0,t}$ arbitrarily. Now, we can integrate this solution, to obtain a trial surface density as $\Sigma_{\rm i,t} = 2 \times \int_{\rm 0}^{\infty} \rho_{\rm i,t}(z) dz$. The surface density is an observable quantity at every radius and can be compared with the trial one ($\Sigma_{\rm i,t}$) to constrain the midplane density. We iteratively update the $\rho_{\rm i,0,t}$, such as it produces a $\Sigma_{\rm i,t}$ close to the observation. We stop this iterative process when the trial surface density matches the observed one with better than 0.1\% accuracy. For our sample galaxies, we find that the surface densities for any component converge within 10s of iterations at any radius. It should be mentioned here that we assumed the observed column density at any radius is fully determined by the vertical density distribution at that radius. For an \HI~disc with finite thickness, this will only be true for a face-on orientation. In any other inclination, the density distribution of the adjacent radii would contribute to the surface density. However, the error introduced due to this effect is expected to be smaller as compared to the observational uncertainties.

%It should be mentioned here that, in earlier studies, selecting a proper midplane density was done manually, i.e., the trial midplane density in the next iteration was updated by looking at how close is the $\Sigma_{i,t}$ to the observed value in the previous iteration. This manual selection might require a large effort to solve Eq.~\ref{eq5} for three components at multiple radial points for many galaxies. For example, to solve Eq.~\ref{eq5} for NGC 2976, the smallest galaxy in our sample, we solve the hydrostatic equation every 100 pc from $0.2 \ kpc \leq R \leq 5.3 \ kpc $, which results in 50 radial points. Solving for three components leads to a total 150 times, the individual differential equation to be solved. This is significant considering the fact that, at each time, at least a couple of iterations is expected to achieve better than 0.1\% convergence accuracy. 

\begin{figure*}
\begin{center}
\begin{tabular}{c}
\resizebox{1.\textwidth}{!}{\includegraphics{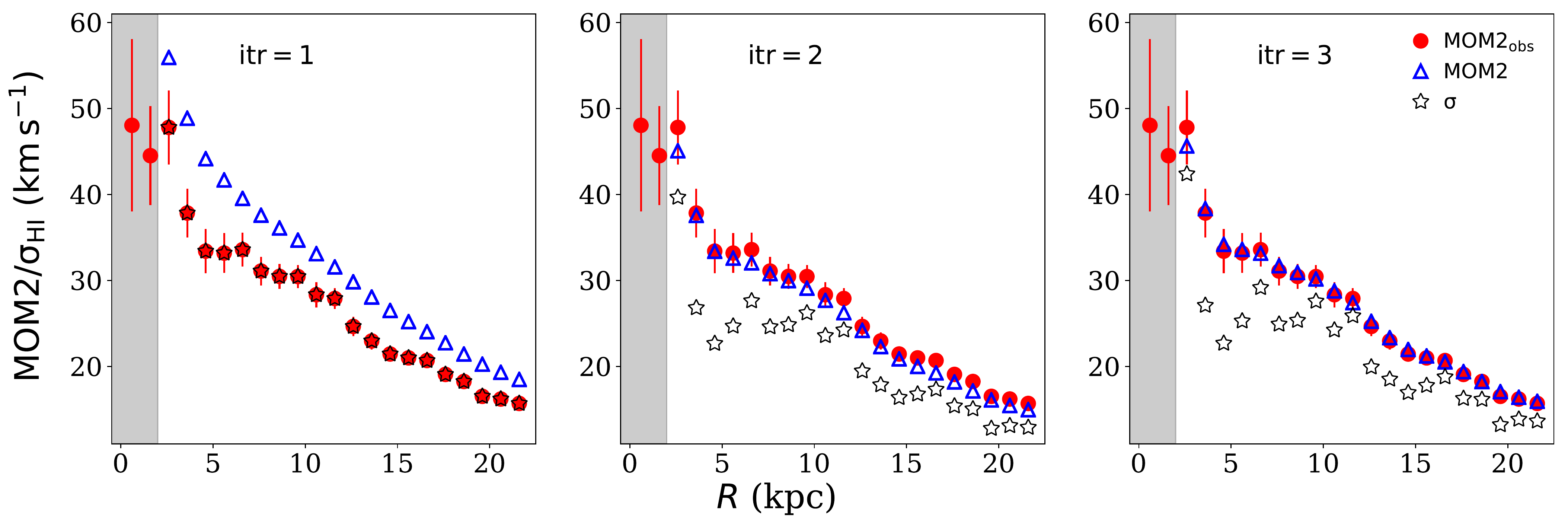}}
\end{tabular}
\end{center}
\caption{The $\sigma$/MOM2 profile of NGC 7331 at different iterations of the iterative method. The solid red circles with error bars in each panel represent the observed MOM2 profile. The empty black stars represent the input $\sigma_{\rm HI}$ profile, whereas the empty blue triangles represent the corresponding simulated MOM2 profiles. The iteration numbers every panel represents are quoted on the top left of every panel. As can be seen, the iterative method gradually finds out a $\sigma_{\rm HI}$ profile, which produces a MOM2 profile much close to the observed one. The shaded regions in each panel represent the radius where the iterative method is not applied due to high spectral blending. See the text for more details.}
\label{sig_itr}
\end{figure*}

%To overcome this limitation, we have developed an automated method to estimate the correct $\rho_{i,0,t}$. We employ a bisection approach to reach the correct midplane density, starting from a coarse guess. We first select a range of the midplane density by trial and error such that, upon solving, the trial surface densities enclose the observed surface densities at all radii. Then in every iteration, we use a bisection method to narrow down the midplane density, which produces a $\Sigma_{i,t}$ closer to the observed one. We continue this bisection method until we hit a $\rho_{i,0,t}$ which produces a $\Sigma_{i,t}$ within an 0.1\% accuracy of the observed one. The rate of convergence of the bisection method depends on the initial trial range; however, for our sample galaxies, this method found to converge within a few 10s of iterations. Thus our code is fully automated and does not require any manual intervention. As the solving of Eq.~\ref{eq5} does not depend on the solutions at any other radius, it can be solved in parallel at multiple radii. We implement our code using MPI based parallel coding and solve Eq.~\ref{eq5} simultaneously at many radial points. It should be emphasized here that the automation of the midplane density search using the bisection method is critical for the parallel implementation of the code.

Eq.~\ref{eq5} represents three coupled differential equations (for three baryonic discs), and hence, in principle, should be solved simultaneously. This is not possible as the exact mathematical form of the coupling between the components is not known. Hence, we adopt an iterative approach to introduce the coupling gradually while solving the equations individually. At any radius, we first solve the three equations (for three components) separately, considering no coupling. For example, in the first iteration, in the gravity term (first term in the RHS of Eq.~\ref{eq5}), we make $\rho_{\rm i}=0$ for all the components except the one we are solving for. In the next iteration onwards, we introduce the gravity of the components through the coupling term. For example, in the first iteration, when we solve for stars, we consider $\rho_{\rm HI} = \rho_{\rm H2} = 0$, i.e., no coupling. Thus we generate a primary solution $\rho_{\rm s,1}(z)$ for the stars. The second index in the subscript represents the iteration number. Similarly, we generate solutions for \HI~and \hh, $\rho_{\rm HI,1}$ and $\rho_{\rm H2,1}$. In the next iteration, we introduce $\rho_{\rm HI,1}$ and $\rho_{\rm H2,1}$ in the gravity term while solving for stars and generate a solution $\rho_{\rm s,2}(z)$. Now, this is a better solution than what is obtained in the first iteration as it includes the gravity of \HI~and \hh~while solving for stars. Similarly, we obtain $\rho_{\rm HI,2}$ and $\rho_{\rm H2,2}$ at the end of the second iteration. However, as $\rho_{\rm HI,1}$ and $\rho_{\rm H2,1}$ was not perfect; these solutions are also not perfect but better than what was obtained in the first iteration. The consecutive iterations are thus run with improved coupled solutions. We repeat this process until the solutions converge, i.e., there is no effective difference between $\rho_{\rm i,n}(z)$ and $\rho_{\rm i,n-1}(z)$, where $n$ is the number of iteration. We note that for our sample galaxies, the coupling quickly converges within a few iterations.

%Using the procedure mentioned above, we solve the hydrostatic equation in our sample galaxies. We solve Eq.~\ref{eq5} every 100 pc, which is adequate to sample our galaxies in detail. The median spatial resolution of our sample galaxies is $\sim 500$ pc. We further use the solutions of Eq.~\ref{eq5} to build detailed three-dimensional dynamical models of our galaxies. 

%As mentioned earlier, $\sigma_{HI}$ is one of the critical input parameters which influences the distribution of \HI~directly. However, though, $\sigma_{HI}$ is not a directly measurable quantity, and what we measure through \HI~spectroscopic observations is the MOM2, intensity weighted $\sigma_{HI}$ along a line-of-sight. This MOM2 very often has a significant contribution from the spectral blending due to the rotational velocities of the galaxy. Nevertheless, this MOM2 can be used to estimate the intrinsic $\sigma_{HI}$ profile while solving Eq.~\ref{eq5}.

As mentioned earlier, to solve Eq.~\ref{eq5}; often, the observed MOM2 is used as a proxy for $\sigma_{\rm HI}$. However, MOM2 is always an overestimate of the $\sigma_{\rm HI}$ due to various smearing effects. To overcome this, we employ an iterative method to estimate the intrinsic $\sigma_{\rm HI}$ profile using the observed MOM2 in galaxies. In the first iteration of this method, we assume the observed MOM2 profile ($\rm M2_{\rm obs}$) to be the input \HI~velocity dispersion profile, $\sigma_{\rm HI,1}$ and solve Eq.~\ref{eq5}. Thus we generate the density solutions for \HI, $\rho_{\rm HI}(R,z)$. This $\rho_{\rm HI}(R,z)$ describes a complete three-dimensional distribution of the \HI~in a galaxy. Using this $\rho_{\rm HI}(R,z)$ and the observed rotation curve, we then build a three-dimensional dynamical model of the galaxy. We then incline this three-dimensional dynamical model to the observed inclination and project it into the sky plane to obtain an \HI~spectral cube sampled with the same velocity resolution as of the observation (1.3 \kms~(one galaxy) or 2.6 \kms~(one galaxy) or 5.2 \kms~(five galaxies) for our sample galaxies). This spectral cube further convolved to the telescope beam to mimic the actual observation. The spectral cube then used to produce the moment maps and hence, a simulated MOM2 profile, $\rm M2_{\rm sim,1}$. In an ideal situation, if the $\sigma_{\rm HI,1}$ was chosen perfectly, the $\rm M2_{\rm sim,1}$ would match the observed MOM2 profile, $\rm M2_{\rm obs}$. However, as the MOM2 is an overestimate of the intrinsic $\sigma_{\rm HI}$, $\rm M2_{\rm sim,1} > M2_{\rm obs}$. 

\begin{figure*}
\begin{center}
\begin{tabular}{c}
\resizebox{1.\textwidth}{!}{\includegraphics{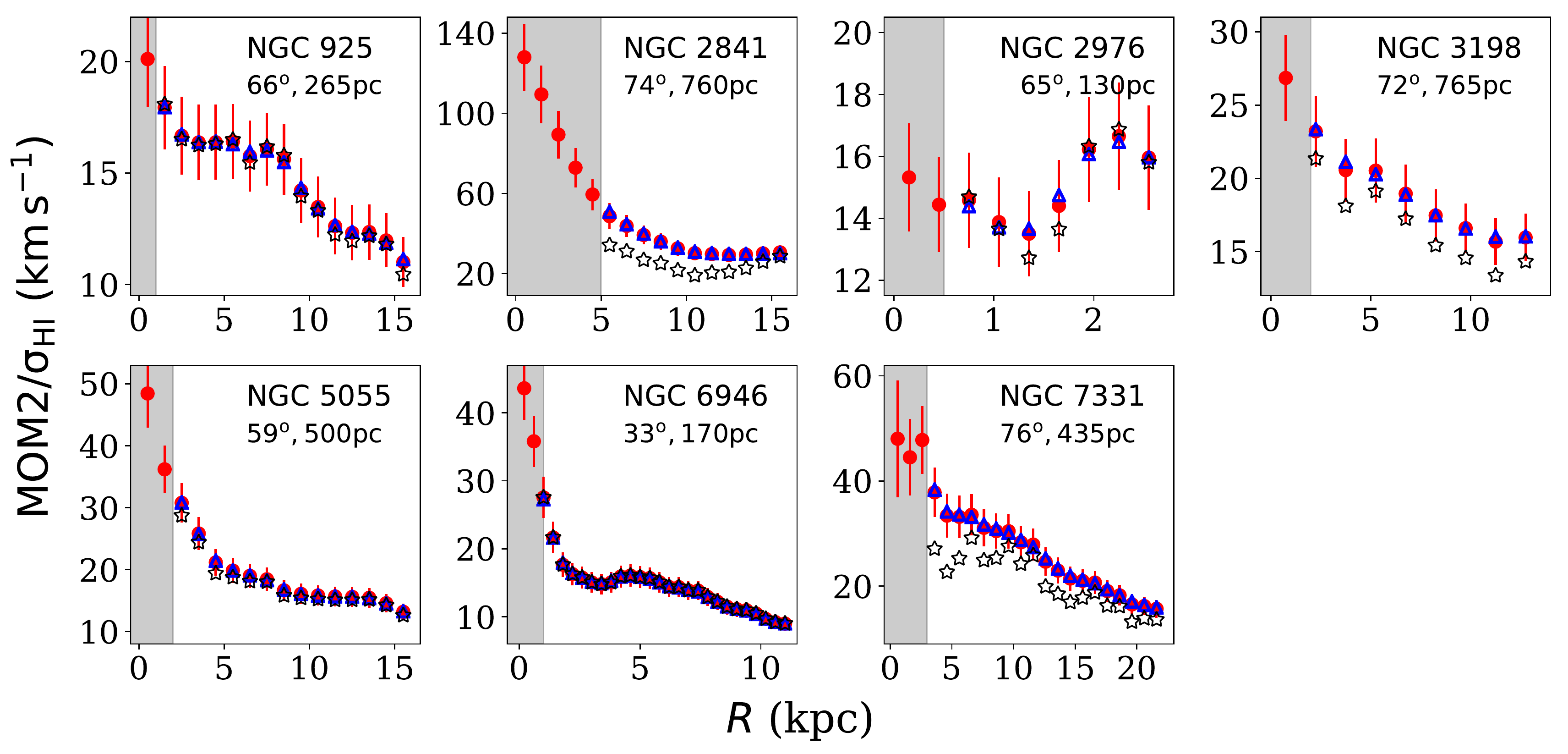}}
\end{tabular}
\end{center}
\caption{The $\sigma_{\rm HI}$/MOM2 profiles of our sample galaxies as extracted by the iterative method. Different panels show the results for different galaxies, as quoted at the top of each panel. The solid red circles with error bars represent the observed MOM2, whereas the empty blue triangles represent the simulated MOM2. The input $\sigma_{\rm HI}$ is shown by the empty black stars. The inclinations and the spatial resolutions of the galaxies are also quoted at the top of every panel. As can be seen, highly inclined galaxies with low spatial resolution suffer maximum spectral blending as indicated by the difference between the recovered $\sigma_{\rm HI}$ and the observed MOM2. See the text for more details.}
\label{sig_opt}
\end{figure*}

\begin{figure*}
\begin{center}
\begin{tabular}{c}
\resizebox{1.\textwidth}{!}{\includegraphics{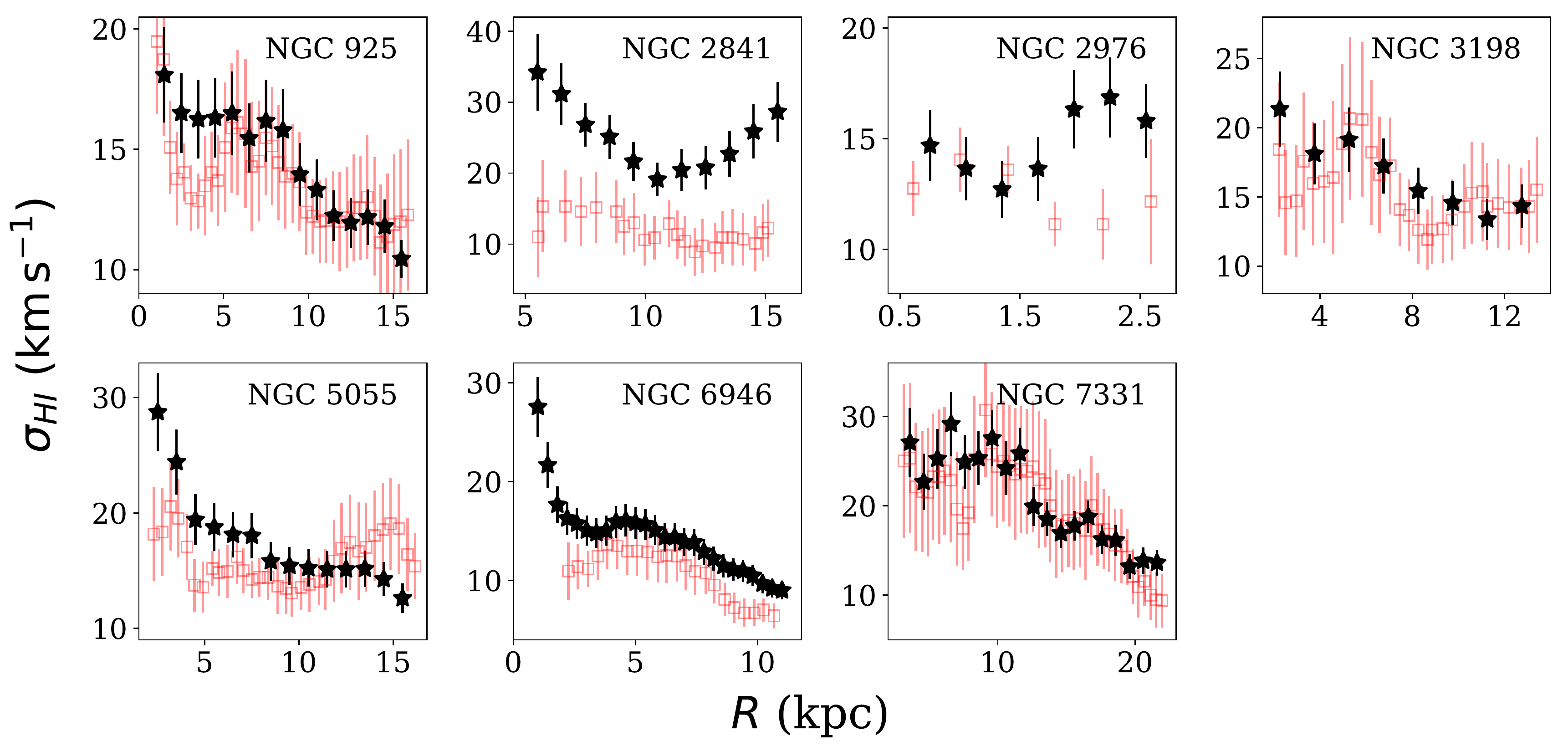}}
\end{tabular}
\end{center}
\caption{Comparison of our $\sigma_{\rm HI}$ profiles to the same as obtained by \citet{bacchini19a}. The solid black asterisks represent the $\sigma_{\rm HI}$ profiles as recovered by the iterative method, whereas the empty red squares indicate the $\sigma_{\rm HI}$ profiles as determined by \citet{bacchini19a}. As can be seen, for most of the galaxies, both the $\sigma_{\rm HI}$ profiles match each other reasonably well. See the text for more details.}
\label{sig_all}
\end{figure*}

Next, we compute the difference between $\rm M2_{\rm sim,1}$ and $\rm M2_{\rm obs}$ at every radius and introduce a proportional change into the input $\sigma_{\rm HI}$ for the next iteration $\sigma_{\rm HI,2}$. Thus, $\sigma_{\rm HI,2}$ is a better estimate of the $\sigma_{\rm HI}$, which will produce a $\rm M2_{\rm sim,2}$ closer to $\rm M2_{\rm obs}$ than what was found in the previous iteration. We continue this process until the difference between the observed and the simulated MOM2 at any radius reduces below a few \kms, which is at least two times better than the velocity resolution of the observation. We note that for our sample galaxies, this method reaches the convergence criteria within a few iterations.

In Fig.~\ref{sig_itr}, we show how the iterative method estimates a $\sigma_{\rm HI}$ profile, which produces the observed MOM2 profile for NGC 7331. In the left panel, we show the first iteration where the observed MOM2 profile (solid red circles with error bars) is taken as the input $\sigma_{\rm HI}$ profile (empty black stars). This produces a simulated MOM2 profile (empty blue triangles), which is higher than the observed one. In the next iteration, we introduce a proportional change in the input $\sigma_{\rm HI}$ profile, as shown in the middle panel. As can be seen, a lower input $\sigma_{\rm HI}$ profile now produces a MOM2 profile, which is closer to the observed one. In the right panel, we show the results from the third iteration. The updated $\sigma_{\rm HI}$ profile in this iteration produces a MOM2 profile that matches the observed ones within our convergence criteria. It should be noted that the satisfaction of the hydrostatic equilibrium condition is crucial for our analysis. Any violation of the same might lead to a wrong interpretation of the results. Observationally, the central regions of spiral galaxies ($\lesssim 1$ kpc) are generally found to exhibit highly energetic activities with associated outflows and enhanced star formation \citep{irwin96,bolatto13}. Subsequently, in these regions, the condition of hydrostatic equilibrium might not be valid fully. Moreover, due to several other effects, e.g., the high gradient in rotation curve, strong non-circular motions, etc., the \HI~spectra in these regions could be broadened by a significant amount. Such as, this artificial broadening dominates the spectral width (especially for high inclination galaxies) as compared to the actual $\sigma_{\rm HI}$. Consequently, in these regions, the effective $\rm M2_{\rm sim}$ becomes insensitive to $\sigma_{\rm HI}$, and the iterative method fails to recover the intrinsic $\sigma_{\rm HI}$ profiles effectively. Further, a high MOM2 value also leads to a divergence of Eq.~\ref{eq5} in the first iteration of the iterative method. Considering these complexities, we exclude a portion of the central regions (by a trial and error method) of our galaxies and do not apply the iterative method. The grey shaded areas in Fig.~\ref{sig_itr} and~\ref{sig_opt}, show the extent of these regions for our galaxies.

\section{Results and discussion}

We apply the above mentioned iterative method in our seven sample galaxies and solve the hydrostatic equilibrium equation. In Fig.~\ref{sig_opt}, we show how the iterative method estimates the $\sigma_{\rm HI}$ profiles in our galaxies using the observed MOM2 profiles. As can be seen from the figure, the iterative method could assess the $\sigma_{\rm HI}$ profiles, such as the output MOM2 profiles matches to the observed ones reasonably well. The inclinations and the spatial resolutions of the observations quoted at the top of every panel in Fig.~\ref{sig_opt}. As can be seen, galaxies with high inclinations and coarse spatial resolutions (e.g., NGC 2841, NGC 3521) exhibit maximum spectral blending, which in turn results in a maximum difference between the observed MOM2 profile and the estimated $\sigma_{\rm HI}$. Galaxies with low inclinations and high spatial resolutions endure negligible or minimal blending. For example, NGC 6946 has an inclination of $33^o$ and was observed with a spatial resolution of $\sim 170$ pc. For this galaxy, the intrinsic $\sigma_{\rm HI}$ profile matches with the observed MOM2 profile well within a few percent. {\it This indicates, for this kind of observational parameters, the observed MOM2 represent the intrinsic $\sigma_{\rm HI}$ profile fairly well.} This has a significant impact on understanding the connection between the velocity dispersion and various physical processes (e.g., star formation, feedback, etc.) in a galaxy.

Using a different approach, very recently, \citet{bacchini19a} had estimated the $\sigma_{\rm HI}$ profiles in a sample of 12 nearby galaxies. They used the \HI~spectral cubes to perform a rigorous 3D tilted-ring modeling using the software $^{3D}$Barolo \citep{diteodoro15} and simultaneously model the \HI~velocity dispersion \citep[see also,][]{iorio17}. They used these $\sigma_{\rm HI}$ profiles to estimate the vertical scale height profiles of the \HI~discs employing a hydrostatic equilibrium condition. In Fig.~\ref{sig_all}, we compare our $\sigma_{\rm HI}$ profiles as estimated by the iterative method (black filled asterisks) to what is determined by \citet{bacchini19a} (empty red squares). As can be seen from the figure, for most of our galaxies, both the $\sigma_{\rm HI}$ profiles match reasonably well. However, for the galaxy NGC 2841, the iterative method recovers much higher $\sigma_{\rm HI}$ values as compared to what is obtained by \citet{bacchini19a}. The inclination of NGC 2841 is $74^o$, and it was observed with a coarse spatial (760 pc) and spectral resolutions (5.2 \kms). Its rotation curve also shows a significant gradient (Fig.~\ref{rotcur}, second panel on the top) across all radius. It is possible that significant smearing is introduced due to these effects, and in these circumstances, the iterative method might not be fully effective in recovering the $\sigma_{\rm HI}$ correctly. We also note that the recovered $\sigma_{\rm HI}$ for the galaxy NGC 7331 is somewhat higher than what is observed for other galaxies in our sample (however, though it matches reasonably well with that of obtained by \citet{bacchini19a}). This could also be due to its high inclination ($76^o$) and a modest spatial resolution ($\sim 450$ pc). We emphasize that despite a completely different approach, our iterative method recovers $\sigma_{\rm HI}$ profiles, which are consistent with those found by \citet{bacchini19a}.

\begin{figure}
\begin{center}
\begin{tabular}{c}
\resizebox{.47\textwidth}{!}{\includegraphics{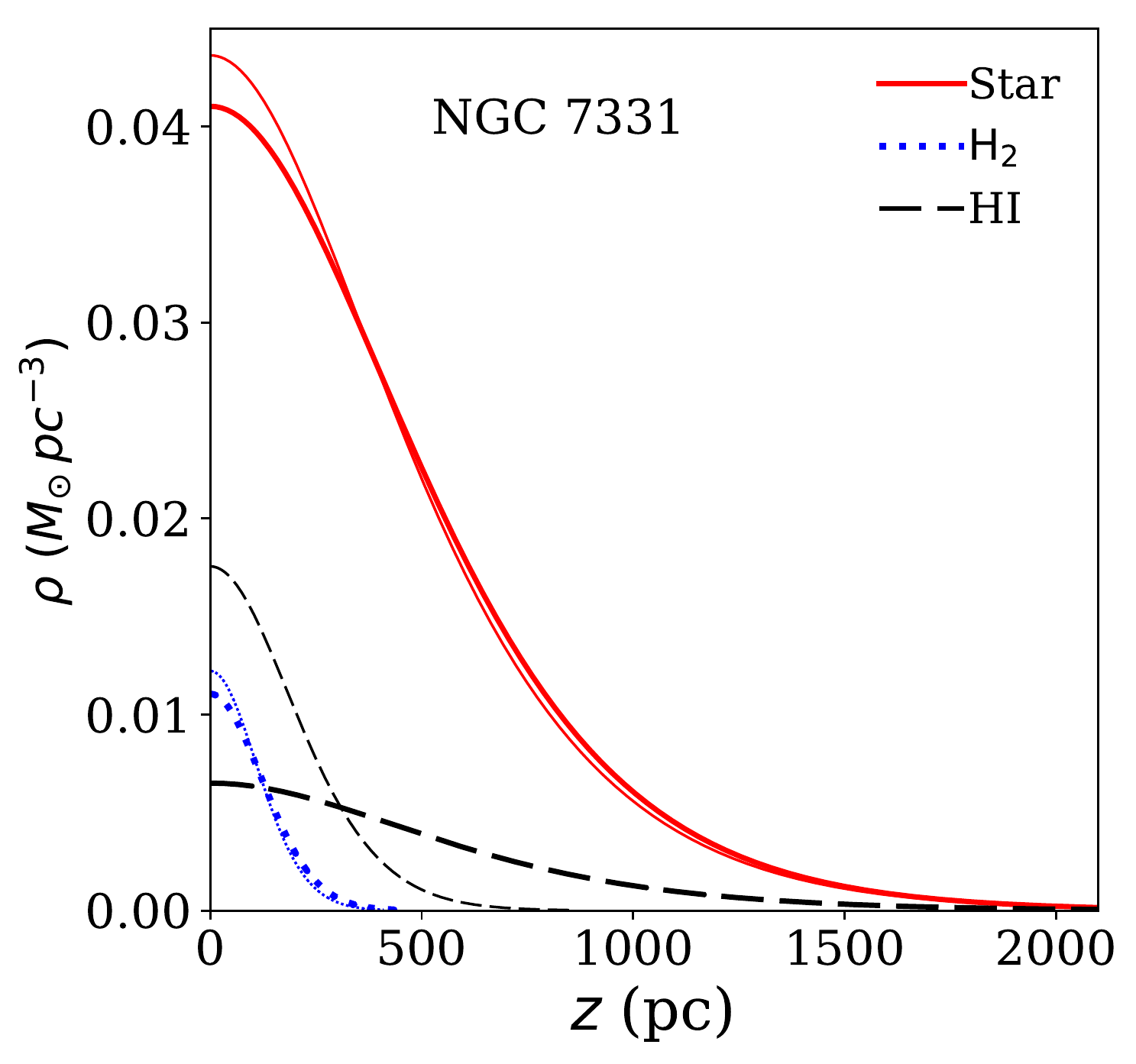}}
\end{tabular}
\end{center}
\caption{The sample solution for NGC 7331 at a radius of 10 kpc. The thick lines represent the solutions using our iterative method (with recovered $\sigma_{\rm HI}$), while the thin lines represent the solutions assuming a constant $\sigma_{\rm HI} = 12$ \kms. The solid red lines describe the volume density of stars, whereas the blue dotted lines represent the density of molecular gas. The black dashed lines denote the solutions for \HI. As can be seen, a constant velocity dispersion for \HI~produces a very different \HI~disc than what would be otherwise. See the text for more details.}
\label{sampl_sol}
\end{figure}

Thus, for all our galaxies, using the iterative method, we self-consistently solve for the density distributions and the $\sigma_{\rm HI}$ profile. In Fig.~\ref{sampl_sol}, we plot a sample solution for NGC 7331 at a radius of 10 kpc. For comparison, we also plot the density solutions for an assumed constant $\sigma_{\rm HI} = 12$ \kms~at the same radius (thin lines). This $\sigma_{\rm HI} = 12$ \kms~is commonly used as the velocity dispersion of the \HI~discs in spiral galaxies as estimated by \citet{calduprimo13} and \citet{mogotsi16} using the THINGS data. As can be seen from the figure, a constant $\sigma_{\rm HI} = 12$ \kms~produces a very different \HI~disc as compared to what is found by the iterative method. The $\sigma_{\rm HI}$ found by the iterative method at a radius of 10 kpc is $\sim$ 28 \kms~which is significantly higher than the constant $\sigma_{\rm HI} = 12$ \kms. This higher $\sigma_{\rm HI}$ significantly flares the \HI~disc (at 10 kpc), resulting in a much larger \HI~scale height. However, though the $\sigma_{\rm HI}$ changes the \HI~density distribution significantly, it marginally influences the \hh~(dotted thin and thick blue lines) and the stellar discs (solid thin and thick red lines), as can be seen from the figure. We note that for a single-component isothermal disc in hydrostatic equilibrium, the density distribution follows a $sech^2$ law \citep{spitzer42}. However, for a galaxy with multiple discs and dark matter halo, the density distribution deviates significantly from $sech^2$ law and resembles more like a Gaussian due to coupling \citep[see, e.g.,][]{olling95,koyama09}.

\begin{figure*}
\begin{center}
\begin{tabular}{c}
\resizebox{1.\textwidth}{!}{\includegraphics{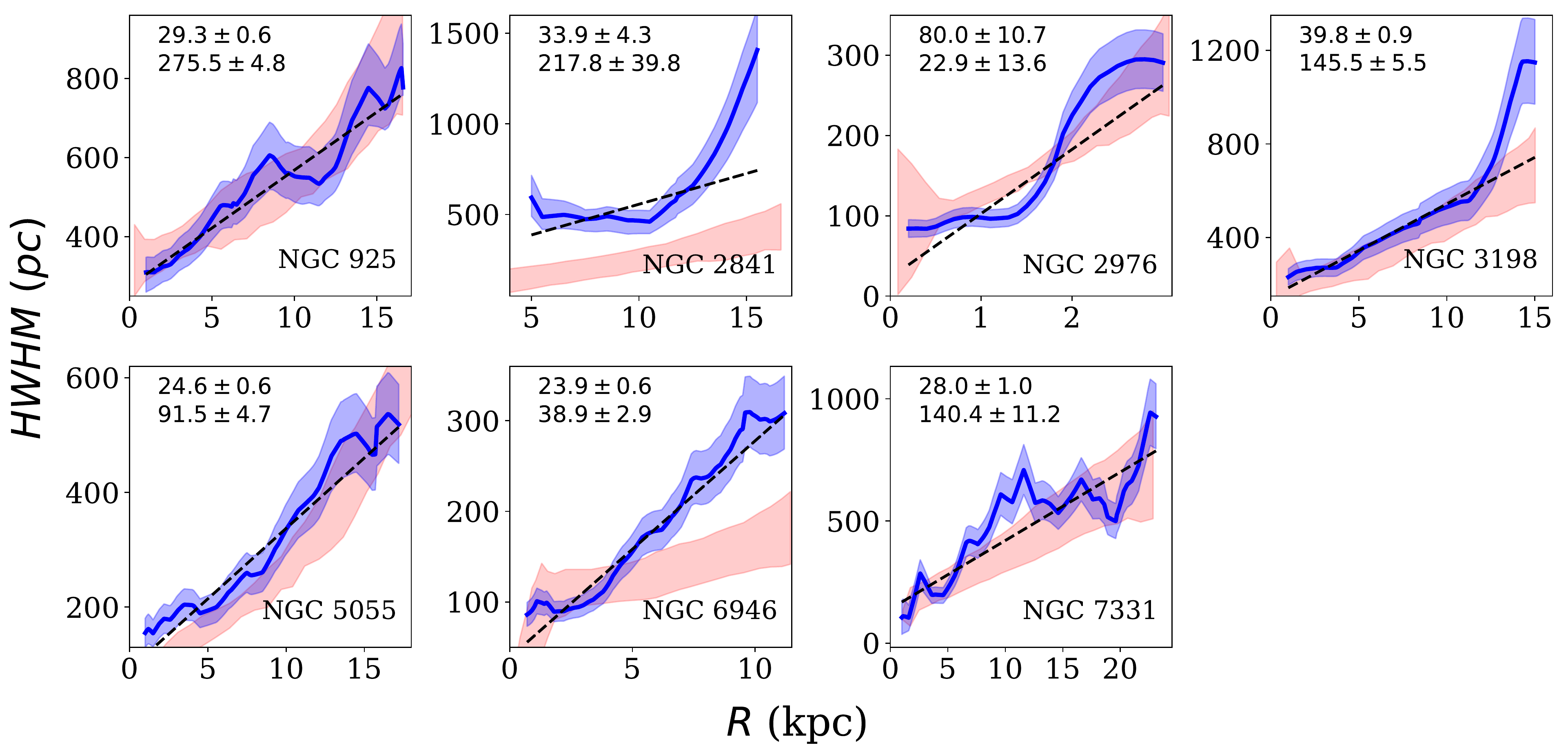}}
\end{tabular}
\end{center}
\caption{The \HI~scale height (HWHM) profiles of our sample galaxies. Different panels represent the scale heights for different galaxies, as quoted in the bottom right corners of every panel. The thick blue lines represent the \HI~scale height profiles, whereas the blue-shaded regions depict its one sigma error bar. The black dashed lines represent a linear fit to these scale height profiles. The two numbers from the top left of every panel indicate the slopes (in the units of pc/kpc) and the intercepts (in the units of pc) of the fitted lines, respectively. The red shaded region in every panel represents the \HI~scale height profiles obtained by \citet{bacchini19a}. As can be seen, for most of the galaxies, our determination of the scale heights compares very well with what is found by \citet{bacchini19a}. See the text for more details.}
\label{sclh}
\end{figure*}

\begin{figure}
\begin{center}
\begin{tabular}{c}
\resizebox{.47\textwidth}{!}{\includegraphics{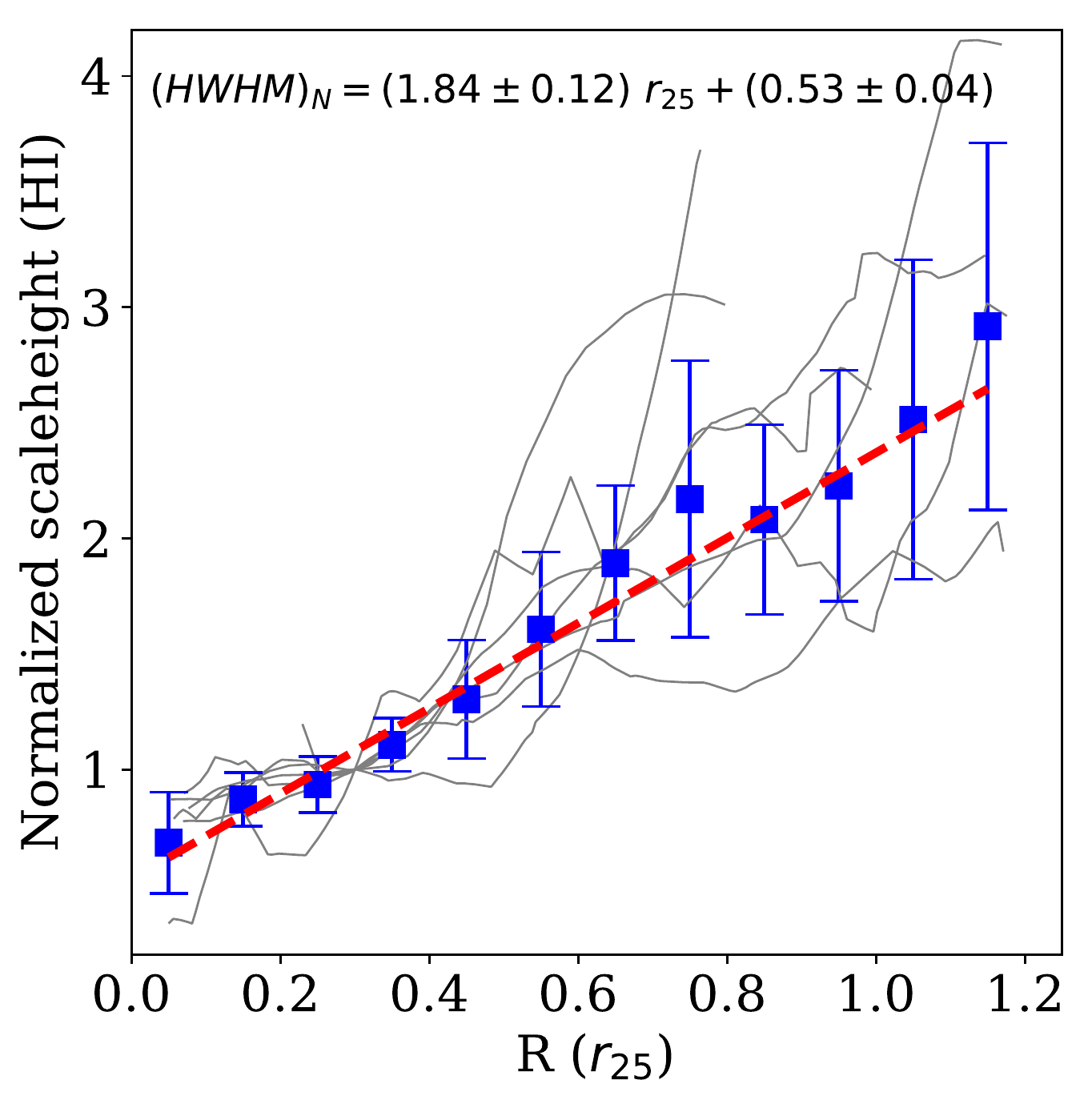}}
\end{tabular}
\end{center}
\caption{The normalized scale height profiles of our sample galaxies. The thin gray lines represent the normalized scale height profiles, whereas the solid blue squares with error bars represent the average normalized scale height values within a bin of width 0.1 $r_{\rm 25}$. The red dashed line represents a linear fit to these average normalized scale heights. The corresponding straight line equation is quoted at the top left of the figure. See the text for more details.}
\label{normed_scl}
\end{figure}

We use the density solutions of our sample galaxies to extract the \HI~scale heights, the Half-Width at Half-Maxima (HWHM). In Fig.~\ref{sclh}, we plot, thus, obtained \HI~scale height profiles (thick blue lines) along with the estimated errors on them (blue shaded region). For comparison, in every panel, we also plot the \HI~scale height profiles obtained by \citet{bacchini19a} (red shaded region). As can be seen, for most of the galaxies, both the \HI~scale height profiles compare well to each other. For NGC 2841 and NGC 6946, however, the scale height profiles differ considerably from what is obtained by \citet{bacchini19a}. This could be attributed to the difference in their $\sigma_{\rm HI}$ profiles, as compared to what is obtained by \citet{bacchini19a} (Fig.~\ref{sig_all}, second and sixth panel). As can also be seen in Fig.~\ref{sclh}, the \HI~discs in our sample galaxies flares as a function of radius. However, the nature of the flaring is not uniform, and the shape of the scale height profiles vary across our sample galaxies. To quantify this flaring further, we fit these \HI~scale height profiles with straight lines (shown by the black dashed lines). We note that the variation of the shape of the scale height profiles across the galaxies is significant, and hence a simple linear fit is employed to describe the profiles reasonably. The respective slope and the intercepts of the fits are quoted at the top left of corners of the individual panels.

To further investigate the global behavior of the \HI~flaring in our sample galaxies, we normalize the \HI~scale height profiles to unity at a radius of 0.3$r_{\rm 25}$, where $r_{\rm 25}$ is the optical radius of a galaxy calculated by fitting 25th magnitude isophote in B-band. \citet{schruba11} had used the same normalizing condition to find a universality in the \hh~surface density profiles in spiral galaxies. Further, \citet{patra19b} used the same condition to normalize the \hh~scale height profiles in spiral galaxies. Their results suggest a global dependence of the molecular surface density/scale height on the characteristic stellar surface density distribution in a galaxy. In Fig.~\ref{normed_scl}, we plot the normalized \HI~scale height profiles (solid gray lines) for our sample galaxies. As can be seen, despite having a substantial scatter in the normalized scale height profiles, they seem to follow a reasonably well linear trend. This indicates that, globally, the normalized scale height profiles follow a universal linear law. To quantify this flaring behavior further, we evaluate a bin average of these normalized scale heights within a radial bin of 0.1 $r_{\rm 25}$ (blue squares with error bars) and fit it with a straight line (red dashed line). We find a slope of 1.84$\pm$0.12 and an intercept of 0.53$\pm$0.04.
We note that our normalized scale height profiles exhibit a significantly higher scatter than what is found by \citet{patra19b} for molecular scale heights. This indicates a weaker global dependence of the \HI~scale height on the stellar surface density profiles.

The \HI~scale height serves as an indicator of the thickness/density distribution in the vertical direction of the \HI~disc. As can be seen in Fig.~\ref{sclh}, the \HI~scale height in our sample galaxies varies between a few hundred parsecs in the central region (within a few kpc) to $\sim 1-2$ kpc at the outskirts of the \HI~discs. This thickness is comparable to that of the Galaxy \citep[see, e.g.,][]{kalberla08,bacchini19b} and is much smaller as compared to what is found in dwarf galaxies by a similar analysis \citep{patra20a}. This indicates that the \HI~discs in spiral galaxies are much thinner as compared to the dwarf galaxies.

\begin{figure}
\begin{center}
\begin{tabular}{c}
\resizebox{.47\textwidth}{!}{\includegraphics{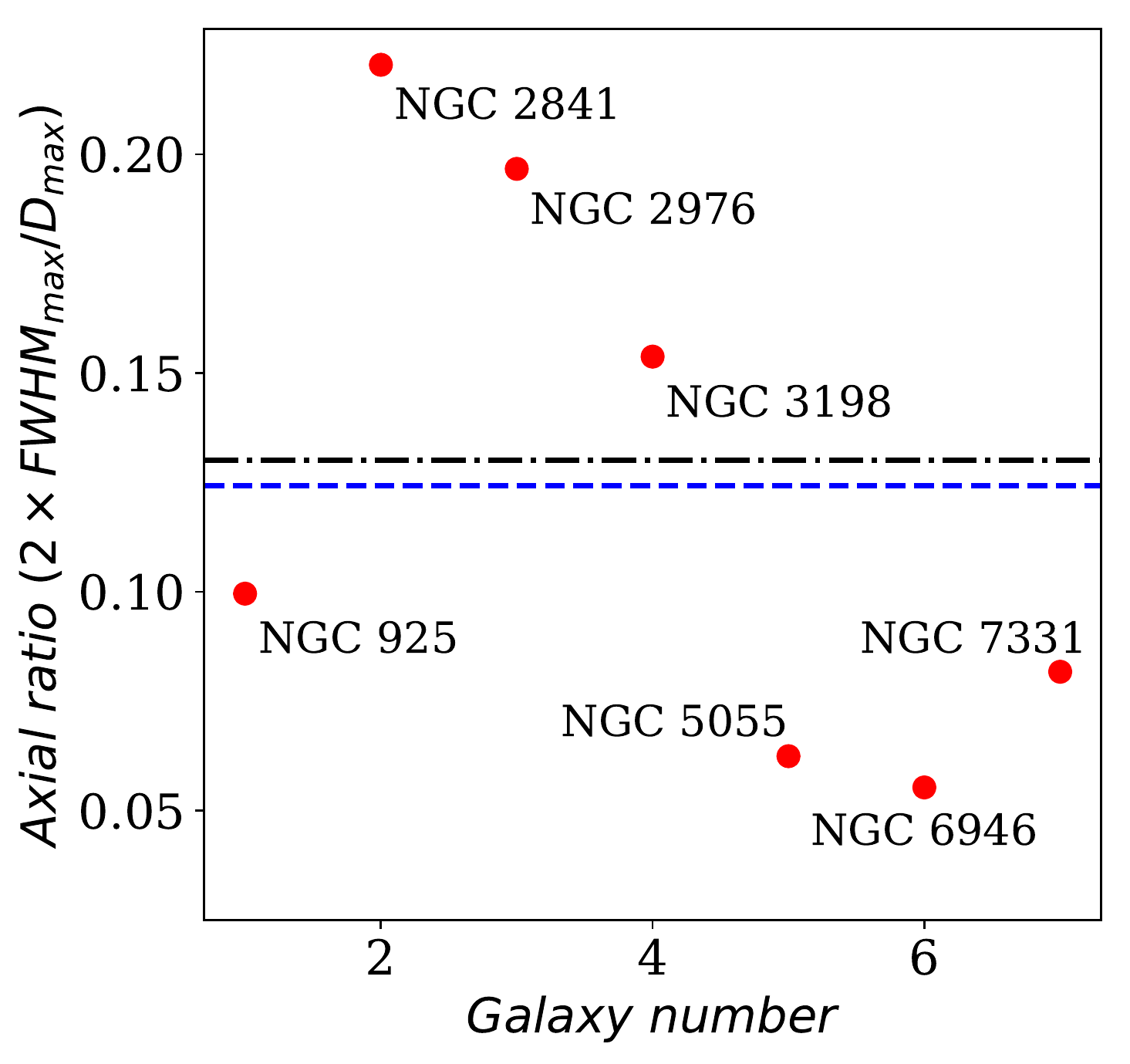}}
\end{tabular}
\end{center}
\caption{The axial ratio of our sample galaxies. Axial ratio is defined as the ratio of twice the FWHM of \HI~distribution to the \HI~diameter of a galaxy. The solid red circles represent the axial ratio for our sample galaxies. The blue dashed line represents the median axial ratio, whereas the black dashed-dotted line represents the same for the Milky Way. As can be seen, the median axial ratio of our sample galaxies matches closely to that of the Galaxy. See the text for more details.}
\label{axial_ratio}
\end{figure}

To make a better sense of the relative thickness of the \HI~discs in our sample galaxies, we estimate the axial ratios using the scale height data. We define the axial ratio as the ratio of the maximum thickness of the \HI~disc (taken to be two times the $\rm FWHM_{\rm max}$) to the diameter of the \HI~disc, $\rm D_{\rm max}$. As FWHM represents only half of the maximum density, we use two times the FWHM (to account for the wings of the Gaussian) as a conservative limit to the maximum \HI~thickness. This ratio is an indicator of the relative thickness of the \HI~discs, which can be compared with other galaxies. In Fig.~\ref{axial_ratio}, we plot the axial ratios of our sample galaxies. The axial ratio varies from $\sim$ 0.06 to $\sim$ 0.22 with a median value of 0.1 (blue dashed line). This median value is very close to the axial ratio found for the Milky Way \citep{kalberla08}, which is $\sim$ 0.13 (black dashed-dotted line). Interestingly, there are three galaxies in our sample, i.e., NGC 5055, NGC 6946, and NGC 7331, for which the axial ratios are much less than 0.1. An axial ratio of $\lesssim 0.1$ for the optical discs in galaxies is traditionally considered an indicator of a superthin galaxy \citep{goad81,karachentsev93,karachentsev99}. However, though there is no clear assessment of the axial ratios of the \HI~discs to qualify a galaxy to be superthin, we adopt the same ratio used for the optical discs as a conservative limit. For example, by performing a detailed 3D modeling, \citet{matthews03} estimated the \HI~scale heights in UGC 7321, a well known superthin galaxy. For this galaxy, the axial ratio of the \HI~disc is found to be 0.07, whereas we find axial ratios of 0.06, 0.06, and 0.08 for NGC 5055, NGC 6946, and NGC 7331, respectively. Mass-modeling of superthin galaxies also revealed that the superthin galaxies host a compact dark matter halo with $R_{\rm C}/R_{\rm D} < 2$, where $R_{\rm C}$ is the core radius of the dark matter halo and $R_{\rm D}$ is the scale length of the optical disc \citep[see, e.g.,][]{banerjee17}. For our three galaxies, $R_{\rm C}/R_{\rm D}$ are found to be 3.7, 1.6, and 1.4 for NGC 5055, NGC 6946, and NGC 7331, respectively. As can be seen, except NGC 5055, the other two galaxies host compact dark matter halos as well. These results suggest that NGC 5055, NGC 6946, and NGC 7331 could possibly represent superthin galaxies. However, detailed comparisons of other physical properties and an edge-on projected view of the baryonic discs would be necessary to make any firm conclusion.

\begin{figure}
\begin{center}
\begin{tabular}{c}
\resizebox{.46\textwidth}{!}{\includegraphics{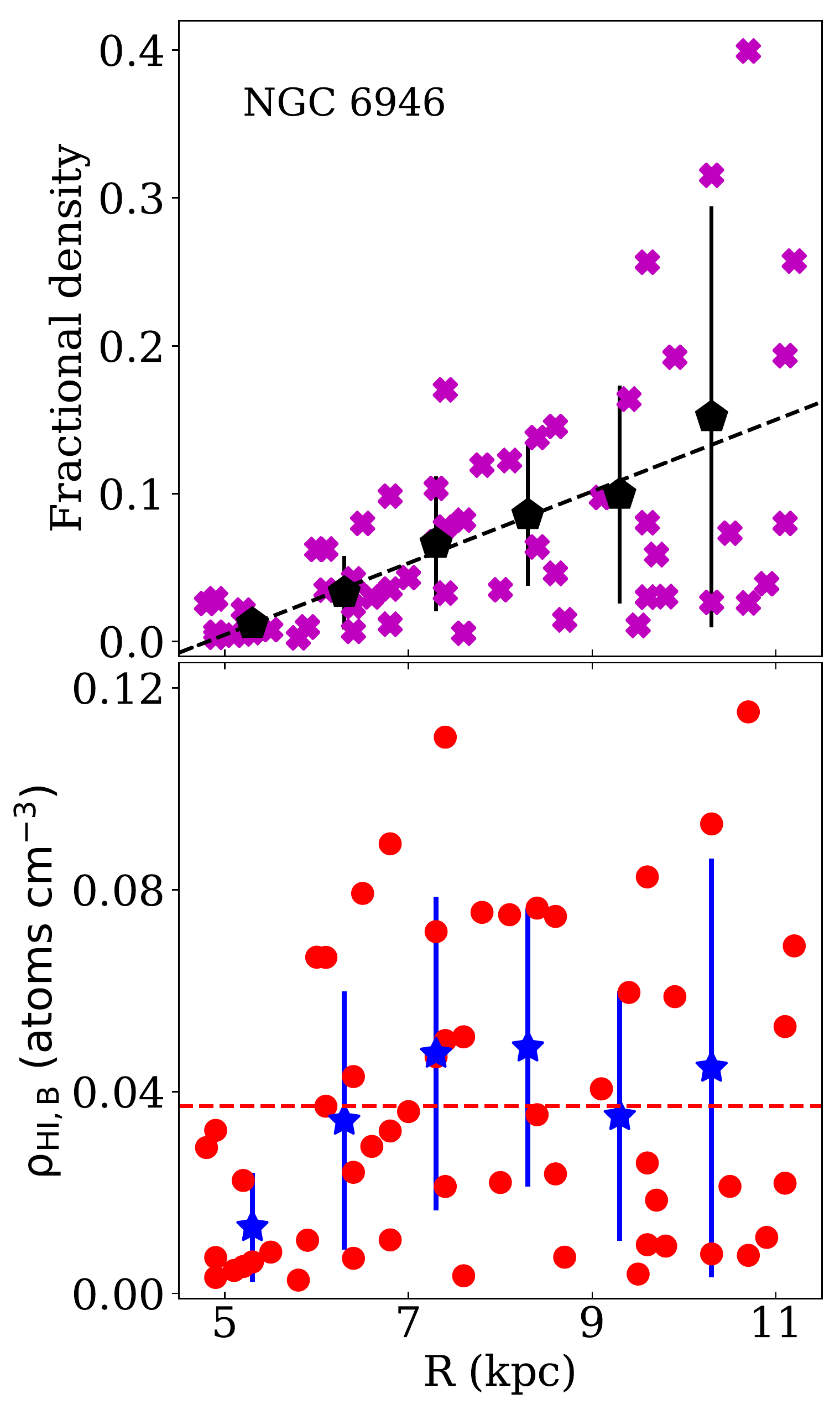}}
\end{tabular}
\end{center}
\caption{Top panel: The fraction of the midplane density at a height equal to the radius of an \HI~hole. The thick magenta crosses represent the fraction, whereas the black pentagons with error bars represent average values within a radial bin of 1 kpc. The black dashed line represents a linear fit to the bin averaged values. See the text for more details. Bottom panel: The solid red circles represent the volume density of the \HI~at a height equal to the radius of an \HI~hole. The blue stars with error bars represent the average values of the same within a radial bin of 1 kpc. The red dashed line represents the mean particle density when an \HI~hole breaks out. See the text for more details.}
\label{hole}
\end{figure}

The thickness of the \HI~discs in galaxies has many significant implications. For example, a thin disc offers a much smaller cross-section to the background quasars to produce \HI~absorption spectra. The thickness of the \HI~discs very often decides the sizes of the \HI~shells or holes forms out of star formation feedback/supernovae. For example, \citet{boomsma08} detected a large number of \HI~holes and high-velocity gas in the typical spiral galaxy NGC 6946 (which is also present in our sample galaxy). They found the average diameter of the \HI~holes to be 1.2 kpc, whereas the diameter of the smallest hole is 766 pc. For NGC 6946, we find that the \HI~scale height reaches up to $\sim 300$ pc at a radius of $\sim$ 12 kpc. This means that the thickness of the \HI~disc in NGC 6946 cannot contain the \HI~holes entirely, and most probably, they have broken out into the immediate environment of the circumgalactic medium (CGM).

To further investigate the effectiveness of the \HI~holes in NGC 6946 in breaking out into the CGM, we calculate the fractional \HI~density at the edge of the \HI~holes in the vertical direction. To do that, we use the size and the location of the detected \HI~holes in NGC 6946 from \citet{boomsma07}. Using our solution at the radius where the \HI~hole is detected, we calculate the volume density of \HI~($\rho_{\rm HI,B}$) at a height equal to the radius of the hole. Using this $\rho_{\rm HI,B}$, and the midplane density at that radius, we determine the density fraction. This can be thought of as the fraction of the midplane density at the edge of the \HI~hole in the vertical direction. In the top panel of Fig.~\ref{hole}, we plot this fraction for all the \HI~holes in NGC 6946 as a function of radius. As can be seen from the figure, for all the holes, the fraction remains less than 0.5, the density point defining the scale height. The majority number of holes have a density fraction of less than 0.2. This means the holes reach much lower densities as compared to the midplane density, strongly indicating a breakout of the \HI~holes into the CGM. As also can be seen from the figure, the density fraction increases as a function of the radius with very low values at the inner galaxies. This indicates that the \HI~holes are more effective in polluting CGM at inner radii as compared to outer radii. To quantify this linear increase, we evaluate an average bin value (black pentagons) withing a radial bin of 1 kpc and fit these average values with a straight line (black dashed line). We find a slope of $0.024 \pm 0.001$ and an intercept of $-0.117 \pm 0.007$ for the linear fit. We would like to mention here that for this analysis, we exclude \HI~holes in a region where we do not solve the hydrostatic equation.

Next, in the bottom panel of Fig.~\ref{hole}, we plot the particle density at the edge of the \HI~holes (in the vertical direction) as a function of radius. As can be seen from the figure, the $\rho_{\rm HI,B}$ vary between $\sim 0.03 - 0.1$ $\rm atoms \thinspace cm^{-3}$ with a mean value of 0.04 $\rm atoms \thinspace cm^{-3}$. We further calculate the average values of this breakout density within radial bins of 1 kpc (shown by the blue filled stars with error bars) and found them to be reasonably constant as a function of radius. However, given the uncertainty, this result is only indicative, and a more detailed study is necessary to draw firm conclusions. 

\section{Conclusion}

In summary, we model galactic discs to be three-component systems consisting of stars, molecular gas, and atomic gas in vertical hydrostatic equilibrium. Under this assumption, we set up the Poisson-Boltzmann equation of hydrostatic equilibrium in a sample of seven large spiral galaxies from the THINGS survey. We solve the hydrostatic equations numerically using an 8$^{\rm th}$ order Runge-Kutta method to estimate the three-dimensional distribution of the \HI~in our sample galaxies.

The \HI~velocity dispersion, $\sigma_{\rm HI}$, is one of the critical inputs which directly influences the vertical distribution of the \HI~in a galaxy. We do not assume the $\sigma_{\rm HI}$ to be constant across our sample or within a galaxy. Instead, we develop an iterative method by which we use the observed MOM2 profile in a galaxy to estimate its intrinsic $\sigma_{\rm HI}$ profile. We find that the iterative method estimates the $\sigma_{\rm HI}$ profiles in our galaxies reasonably well, such as the output MOM2 profiles match the observation within a few percents. Our method self-consistently computes the $\sigma_{\rm HI}$ profile for a galaxy, which is not possible to directly estimate due to line-of-sight integration effects.

Using this iterative method, we solve the hydrostatic equilibrium equation to obtain a detailed three-dimensional distribution of the \HI~in our sample galaxies. Our method reveals that though a constant $\sigma_{\rm HI} = 12$ \kms~does not affect the density distribution of the stars and the molecular gas considerably, it alters the distribution of the \HI~in a significant way. Hence, a precise estimation of the $\sigma_{\rm HI}$ is essential to determine the distribution of \HI~in galaxies accurately. From the solutions of the hydrostatic equation, we also find that due to coupling between the disc components and the presence of the dark matter halo, the density of different disc components deviates from the traditional $sech^2$ law and behaves more like a Gaussian function as also found by previous studies.

Using the density solutions for the \HI~disc, we estimate the \HI~scale heights (Half-Width at Half-Maxima) in our sample galaxies. We find that the \HI~scale height increases as a function of radius, i.e., the \HI~disc flares. This flaring can be represented reasonably well by a linear increase. The scale height of the \HI~discs in our sample galaxies are found to vary between a few hundred parsecs at the center to $\sim 1-2$ kpc at the edge of the \HI~discs. This scale height is much lower as compared to what is observed in dwarf galaxies, indicating thinner \HI~discs in spiral galaxies. 

We further investigate the universality of the \HI~flaring in our galaxies by normalizing the \HI~scale height profiles to unity at a radius of 0.3 $r_{\rm 25}$. Despite a considerable variation in the individual scale height profiles, the normalized profiles are found to follow a linear behavior with a slope of 1.84$\pm$0.12 and an intercept of 0.53$\pm$0.04. The scatter in the normalized \HI~scale height profiles are found to be much higher than what is observed for the molecular scale height profiles. This indicates a weaker global dependence of \HI~scale height on the stellar surface density profiles. 

We quantify the thickness of the \HI~discs in our sample galaxies by estimating an axial ratio, the ratio of two times the maximum FWHM of the \HI~distribution to the \HI~diameter of a galaxy. This axial ratio serves as a measure of the thickness of the \HI~discs in galaxies. We find that the axial ratios for the \HI~discs in our galaxies vary between $\sim$ 0.06 to 0.22 with a median of $\sim$ 0.1. This median axial ratio is very close to what is found for the Milky Way and is much less than what is found for the dwarf galaxies. Using this axial ratio as an indicator, we identify three galaxies in our sample, i.e., NGC 5055, NGC 6946, and NGC 7331 as potential candidates for superthin galaxies.

We use the three-dimensional density distribution of \HI~in one of our sample galaxies, NGC 6946, to estimate the volume density as a fraction of the midplane density at the edge of the \HI~holes (in the vertical direction) as detected by \citet{boomsma08}. We find that most of the \HI~holes reach a density, which is $\lesssim 10-20\%$ of the midplane density. This indicates that most of the \HI~holes in NGC 6946 are broken into the circumgalactic medium (CGM). We also find a linear increase of this fraction as a function of radius, meaning a more efficient metal mixing in the IGM in the inner galaxy as compared to the outer galaxy. We also measure the volume density of \HI~at the edge of the \HI~holes (breakout density) and find a mean value of 0.04 $\rm atoms \thinspace cm^{-3}$. Despite considerable variation in this breakout density amongst the holes, the bin averaged values seem to be reasonably constant as a function of radius. This further strengthens the possibility that the holes are broken out into the CGM. 

Our study points out that it is vital to determine the vertical thickness and the distribution of \HI~accurately in galaxies to understand various physical processes that influence its evolution. Our iterative method offers a simultaneous estimation of the $\sigma_{\rm HI}$ and the \HI~distribution in a self-consistent manner. Given the observing parameters, our method can be used to estimate the amount of spectral blending in a galaxy, which can play a crucial role in interpreting \HI~spectral data.

\section{Acknowledgement}

This work extensively uses data from the THINGS survey. NNP would like to thank the THINGS team for making the data publicly available. NNP would like to thank the anonymous referee gratefully for producing a very detailed, thorough, and thoughtful review of the manuscript. It significantly improved the quality of the paper.

\section{Data availability}
We used already existing publicly available data for this work. No new data were generated or analyzed in support of this research. The scale height and the $\rm \sigma_{HI}$ profiles are available from the author on a reasonable request.

\bibliographystyle{mn2e}
\bibliography{bibliography}

\end{document}